\documentclass[twocolumn,dvipsnames,twocolappendix]{aastex631}
\hypersetup{linkcolor=blue,citecolor=Green,urlcolor=cyan}

\usepackage{amsmath}
\usepackage{textcomp}
\usepackage{gensymb}
\usepackage{graphicx}
\usepackage{tabularx}
\usepackage{url}
\usepackage[caption=false]{subfig}
\usepackage[T1]{fontenc}
\usepackage{float} 
\usepackage{chngcntr}

\newcommand{\noun}[1]{\textsc{\MakeLowercase{#1}}}

\newcommand{\code}{\texttt}

\newcommand{\xmm}{XMM-Newton}
\newcommand{\xmmom}{XMM-OM}
\newcommand{\gaia}{Gaia}
\newcommand{\galex}{GALEX}
\newcommand{\ultrasat}{ULTRASAT}
\newcommand{\gphoton}{g\noun{photon}}
\renewcommand{\O}[1]{O\textsubscript{#1}}
\renewcommand{\H}[1]{H\textsubscript{#1}}
\newcommand{\N}[1]{N\textsubscript{#1}}

\shorttitle{NUV M-dwarf flares with GALEX}
\shortauthors{Rekhi et al.}

\begin{document}

\title{A Census of NUV M-Dwarf Flares Using Archival GALEX Data and the gPhoton2 Pipeline}
\author[0009-0001-3501-7852]{Param Rekhi}
\affiliation{Department of Particle Physics and Astrophysics, Weizmann Institute of Science, Rehovot 7610001, Israel}
\author[0000-0001-6760-3074]{Sagi Ben-Ami}
\affiliation{Department of Particle Physics and Astrophysics, Weizmann Institute of Science, Rehovot 7610001, Israel}
\author[0000-0002-6859-0882]{Volker Perdelwitz}
\affiliation{Department of Earth and Planetary Sciences, Weizmann Institute of Science, Rehovot 7610001, Israel}
\author[0000-0003-1525-5041]{Yossi Shvartzvald}
\affiliation{Department of Particle Physics and Astrophysics, Weizmann Institute of Science, Rehovot 7610001, Israel}



\begin{abstract}
M-dwarfs are common stellar hosts of habitable-zone exoplanets. NUV radiation can severely impact the atmospheric and surface conditions of such planets, making characterization of NUV flaring activity a key aspect in determining habitability. We use archival data from the \galex{} and \xmm{} telescopes to study the flaring activity of M-dwarfs in the NUV. The \galex{} observations form the most extensive dataset of M-dwarfs in the NUV to date, with exploitation of this data possible due to the new \gphoton2 pipeline. We run a dedicated algorithm to detect flares in the pipeline produced lightcurves and find some of the most energetic flares observed to date within the NUV bandpass, with energies of $\sim10^{34}$ ergs. Using GALEX data, we constrain flare frequency distributions for stars from M0 to M6 in the NUV up to $10^5\,$s in equivalent duration and $10^{34}$ ergs in energy, orders of magnitude above any previous study in the UV.
We speculate the combined effect of NUV luminosities and flare rates of M4 and later stars could potentially allow abiogenesis on their habitable zone exoplanets. As a counterpoint, we estimate that the high frequencies of energetic UV flares and associated coronal mass ejections would inhibit the formation of an ozone layer, possibly preventing genesis of complex Earth-like lifeforms due to sterilizing levels of surface UV radiation.
We also provide a framework for future observations of M-dwarfs with \ultrasat{}, a wide FoV NUV telescope to be launched in 2027.
\end{abstract}

\section{Introduction}

M-dwarfs are of high interest as hosts of habitable planets. They comprise approximately 70\% of the galactic stellar population \citep{covey_luminosity_2008, bochanski_luminosity_2010} and are are extremely long-lived on the main sequence \citep{laughlin_end_1997}, offering long timescales for the genesis and evolution of life. These stellar hosts offer several observational advantages as well.
Habitable zone Earth-size planets are easier to detect around smaller stars with radial velocity and transit techniques due to favorable planet-star mass and radius ratios. The lower luminosities and temperatures of M-dwarfs cause their habitable zone to be closer to the host star, increasing the probability of observing a transit \citep{gould_sensitivity_2003, nutzman_design_2008}. Finally, they are well-suited for atmospheric characterization via transit spectroscopy by telescopes such as JWST \citep{kaltenegger_transits_2009, rauer_potential_2011, lustig-yaeger_jwst_2023}. For a detailed discussion of M-dwarfs as habitable exoplanet hosts, see \cite{shields_habitability_2016}.

One of the main drivers of atmosphere evolution and habitability is the UV radiation around stellar hosts. M-dwarfs are known to exhibit high levels of activity in the UV band, flaring at rates and relative amplitudes orders of magnitudes higher than those known for other main sequence stars \citep{gunther_stellar_2020, althukair_main-sequence_2022}, and exhibiting large variation in quiescent UV luminosites within individual spectral types \citep{miles_hazmat_2017, schneider_hazmat_2018, cifuentes_carmenes_2020}. 
Due to their smaller size, M-dwarfs have mostly convective interiors, transitioning to fully convective at spectral type M3-M4 \citep{chabrier_structure_1997, baraffe_closer_2018}. As the stellar magnetic dynamo is linked to the convection layer, and the tachocline between the stellar convection and radiative zones, M-dwarfs are speculated to have a different magnetic dynamo, which might result in higher activity \citep{hall_stellar_2008, charbonneau_dynamo_2010, fan_magnetic_2021}. 
High UV fluxes due to flares can drastically impact the composition of an atmosphere, significantly altering concentrations of species such as \O2, \O3, \H2O, C\O2, C\H3X and \N2O \citep{grenfell_response_2012, venot_influence_2016, airapetian_prebiotic_2016}. Different levels of near-UV (NUV, 190-270 nm) and far-UV (FUV, 140-170 nm) fluxes can either wipe out \O3 from the atmosphere, thereby potentially exposing lifeforms to lethal levels of UV radiation \citep{segura_ozone_2003, tilley_modeling_2019}, or build up \O2 and \O3 to levels that can misinterpreted as false-positive bio-signatures (\citealt{tian_history_2015, harman_abiotic_2015, schaefer_predictions_2016, meadows_exoplanet_2018}; See \citealt{schwieterman_exoplanet_2018} for a recent discussion of bio-signatures of exoplanets, and \citealt{segura_biosignatures_2005, scalo_m_2007, grenfell_sensitivity_2014, ranjan_photochemical_2022} for a discussion specific to M-dwarfs).
Coronal mass ejections (CMEs) accompanying flares can also significantly impact atmospheric chemistry \citep{tilley_modeling_2019}. High X-ray and extreme UV emissions (X-ray, <10 nm; EUV, 10–120 nm; together XUV), along with CMEs, can strip atmospheres, rendering planets uninhabitable \citep{lammer_coronal_2007, luger_habitable_2015}, while lower levels of XUV radiation may wipe out the water content of habitable zone exoplanets during the pre-main-sequence stage of their host M star \citep{tian_history_2015, tian_water_2015}.

While the discussion above emphasizes UV radiation as a stressor for life, a certain level of NUV radiation is thought to be a necessary component for the origin of life on Earth \citep[abiogenesis;][]{ranjan_influence_2016}. As UV radiation is thought to be an essential energy source for prebiotic synthesis of compounds leading to abiogenesis (such as mRNA), the low quiescent UV radiation of M-dwarfs entails the importance of flares \citep{ranjan_surface_2017, rimmer_origin_2018}. \cite{mullan_photosynthesis_2018} further argue that M-dwarf flares may also provide a source for otherwise scarce visible-light photosynthesis.

In the past there has been a paucity of information on the frequency and energy distribution of UV flares of M-dwarfs due to their relatively low brightness and the long observational spans required to satisfactorily fill the gap in our knowledge. 
Over the last decade, there have been big strides in UV photometric and spectroscopic observations of M-dwarfs by the MUSCLES and HAZMAT projects \citep{france_muscles_2016, shkolnik_hazmat_2014}, based primarily on HST data. \cite{loyd_muscles_2018} and \cite{parke_loyd_hazmat_2018} showed that M-dwarf flare rates in the FUV as a function of equivalent duration are agnostic to age and optical activity markers. 
Studies of M-dwarfs in the NUV have been undertaken previously, but were either focused on quiescent trends \citep{miles_hazmat_2017, schneider_hazmat_2018}, or studied flares with a limited dataset of M-dwarfs and combined observation spans of less than a day.
\cite{hawley_nearultraviolet_2007, kowalski_near-ultraviolet_2019, chavali_pilot_2022} conducted spectroscopic observations of M-dwarf flares in the NUV. While limited in sample size, these studies have obtained important results regarding the relative contributions of continuum and line emissions in NUV flare spectra. Limited flare studies with \galex{} have been conducted by \cite{welsh_detection_2007}, who searched for flaring M-dwarfs in early \galex{} observations, \cite{fleming_new_2022}, who studied flares in the GJ 65 system and \cite{brasseur_short-duration_2019, brasseur_constraints_2023}, who conducted a large survey of flares in \galex{} and Kepler data, albeit focused on G-stars with a limited sample of M-dwarfs. There have also been multi-wavelength flare studies in the NUV/optical and NUV/X-ray regimes conducted by \cite{hawley_nearultraviolet_2007, paudel_simultaneous_2021, jackman_extending_2022} and \cite{jackman_characterization_2022}.

There have been multiple extensive studies of M-dwarf flares in the visible/NIR bands (white-light flares) with the Kepler and TESS telescopes \citep{gunther_stellar_2020, jackman_stellar_2021, stelzer_flares_2022, althukair_main-sequence_2022}, but these are of limited applicability to the question of exoplanetary habitability, and extrapolation of white-light flare rates to the UV have been shown to be prone to significant errors \citep{kowalski_near-ultraviolet_2019, jackman_extending_2022}.

In this work we present M-dwarf flare observations in the NUV carried out with \galex{} and the \xmm{} Optical Monitor, with the \galex{} dataset being the largest sample of M-dwarfs yet studied for flaring activity, both in number of stars as well as observational durations. The \xmmom{} sample proves to be too small for statistical analysis, yet we present it here as a demonstration of its capabilities to detect and characterize M-dwarf flares at high cadences. In section \ref{sec: Dataset}, we describe the stellar sample, observations and data reduction. In section \ref{sec: flare characterization}, we describe the flare identification process and characterization of the detected flares. Section \ref{sec: Data Analysis} exhibits the NUV flare frequency distributions, as well as a long baseline activity study. We discuss our results in section \ref{sec: Discussion}, with special emphasis on their implications on exoplanetary habitability. A framework for future observations of M-dwarfs with \ultrasat{}, a wide FoV NUV telescope to be launched in 2027 is given in section \ref{sec: ultrasat}. A summary of our work appears in section \ref{sec: summary}. 

\section{Data} \label{sec: Dataset}

\begin{figure*}
    \centering
    \subfloat[]{
        \includegraphics[width=\columnwidth]{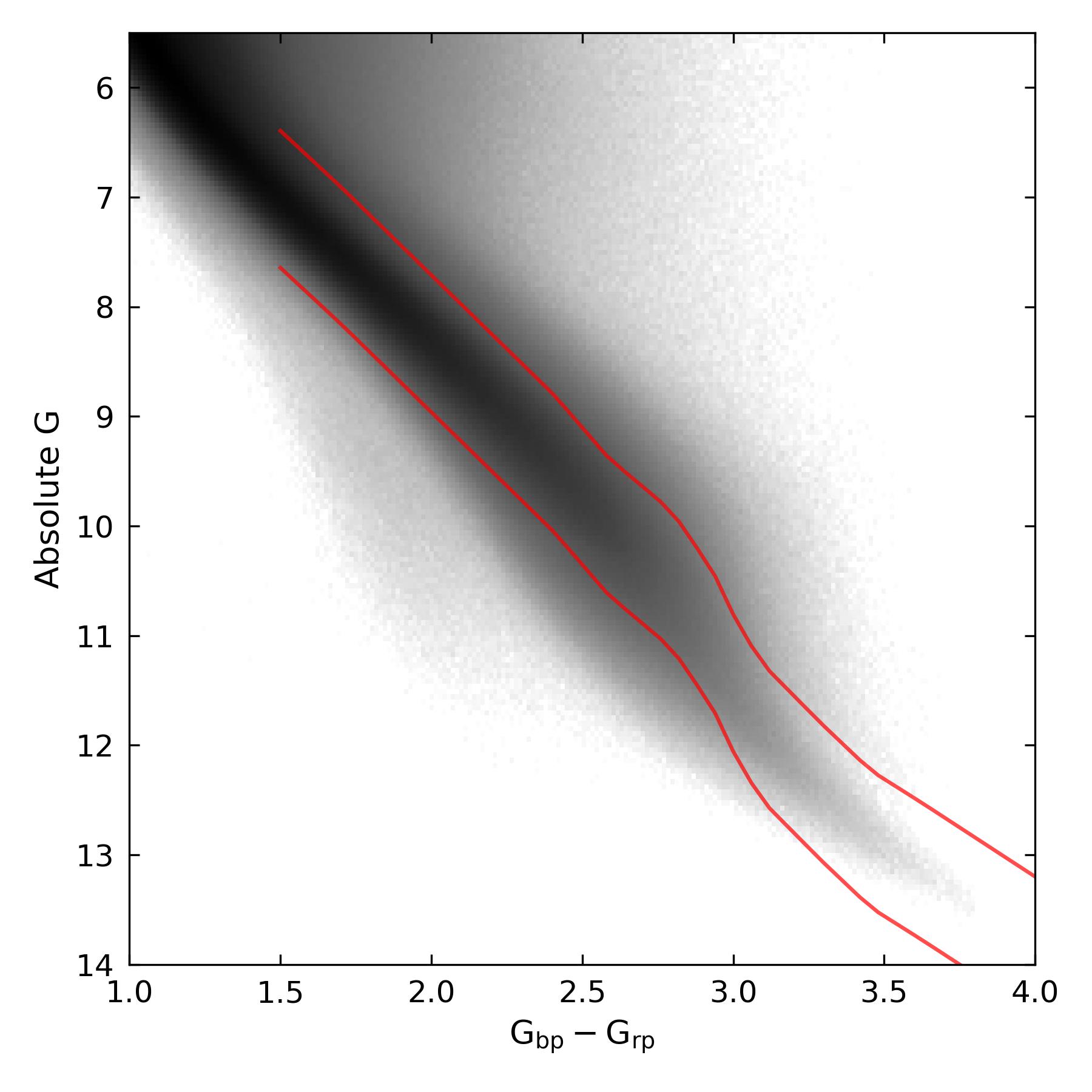}
        \label{subfig: Gaia DR3 HRD}}
    \subfloat[]{
        \includegraphics[width=\columnwidth]{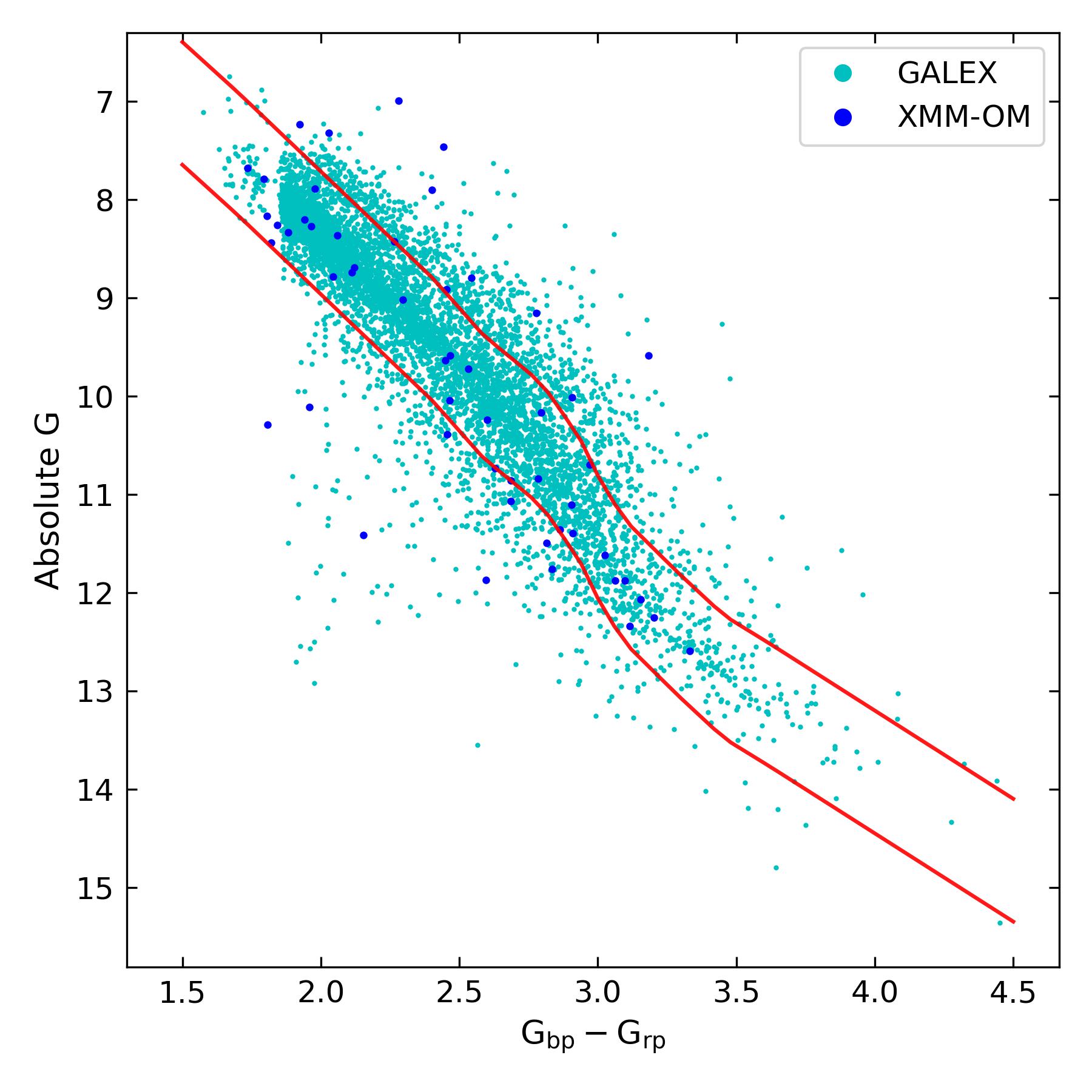}
        \label{subfig: sample HRD}}
    \caption{Gaia DR3 color magnitude diagram (CMD)  plotted for stars from the (a) Gaia DR3 survey and (b) \galex{} and \xmmom{} cross-matched samples. The red curves demarcate the main sequence bounds $+0.65<M_G<-0.6$ described in section \ref{sec: Dataset}.}
    \label{fig: HRD}
\end{figure*}

Our analysis is conducted on archival datasets from the Galaxy Evolution Explorer (GALEX) and the \xmm{}  Optical Monitor (XMM-OM). The \galex{} lightcurves are generated using the \gphoton2  pipeline \citep{st_clair_gphoton2_2022}, while the \xmmom{} data are available in the form of pre-generated lightcurves downloadable from the \xmm{} Science Archive\footnote{\url{http://nxsa.esac.esa.int}} (XSA). We detail the methodology of compiling these datasets, and their scope, in the following sections.

We use the TESS Input Catalog v8.2 (TIC) \citep{stassun_revised_2019, paegert_tess_2021} to generate a list of M-dwarfs comprising of objects with TIC effective Temperature (\code{Teff}) below 3900K and luminosity class (\code{LClass}) \noun{`DWARF'}, within 200\,pc. This results in a list of 1.22 million stars from which we identify targets in \galex{} and \xmm\ data by a cross-match after propagating the Gaia DR2 positions (epoch J2015.5) \citep{brown_gaia_2018} of the TIC M-dwarfs to the epochs of the observations. We categorize the matched M-dwarfs into their spectral sub-types using their TIC temperatures and the color-temperature relations given by \cite{pecaut_intrinsic_2013}\footnote{\url{http://www.pas.rochester.edu/~emamajek/EEM_dwarf_UBVIJHK_colors_Teff.txt}}. We obtain the most current geometric distance estimates to these sources from \cite{bailer-jones_estimating_2021} based on astrometry from Gaia DR3.

Figure \ref{subfig: Gaia DR3 HRD} shows a color-magnitude diagram constructed with Gaia DR3 BP-RP colors and G-band absolute magnitudes \citep{vallenari_gaia_2022}, using the query parameters described in appendix B of  \cite{babusiaux_gaia_2018} for quality control. We empirically demarcate the main sequence as the region bounded by $\Delta M_G=+0.65$ and $\Delta M_G=-0.6$ around the maximum probability curve. The lower bound removes the majority of equal-mass binaries and pre-main-sequence stars, while the upper bound excludes cool subdwarfs. We base these cuts on previous work by \cite{jackman_stellar_2021}, adjusted by visual inspection to the Gaia color-magnitude diagram. We exclude the cross-matched sources falling outside these bounds as shown in Figure \ref{subfig: sample HRD}.

\begin{figure*}
    \centering
    \subfloat[]{
    \includegraphics[width=\columnwidth]{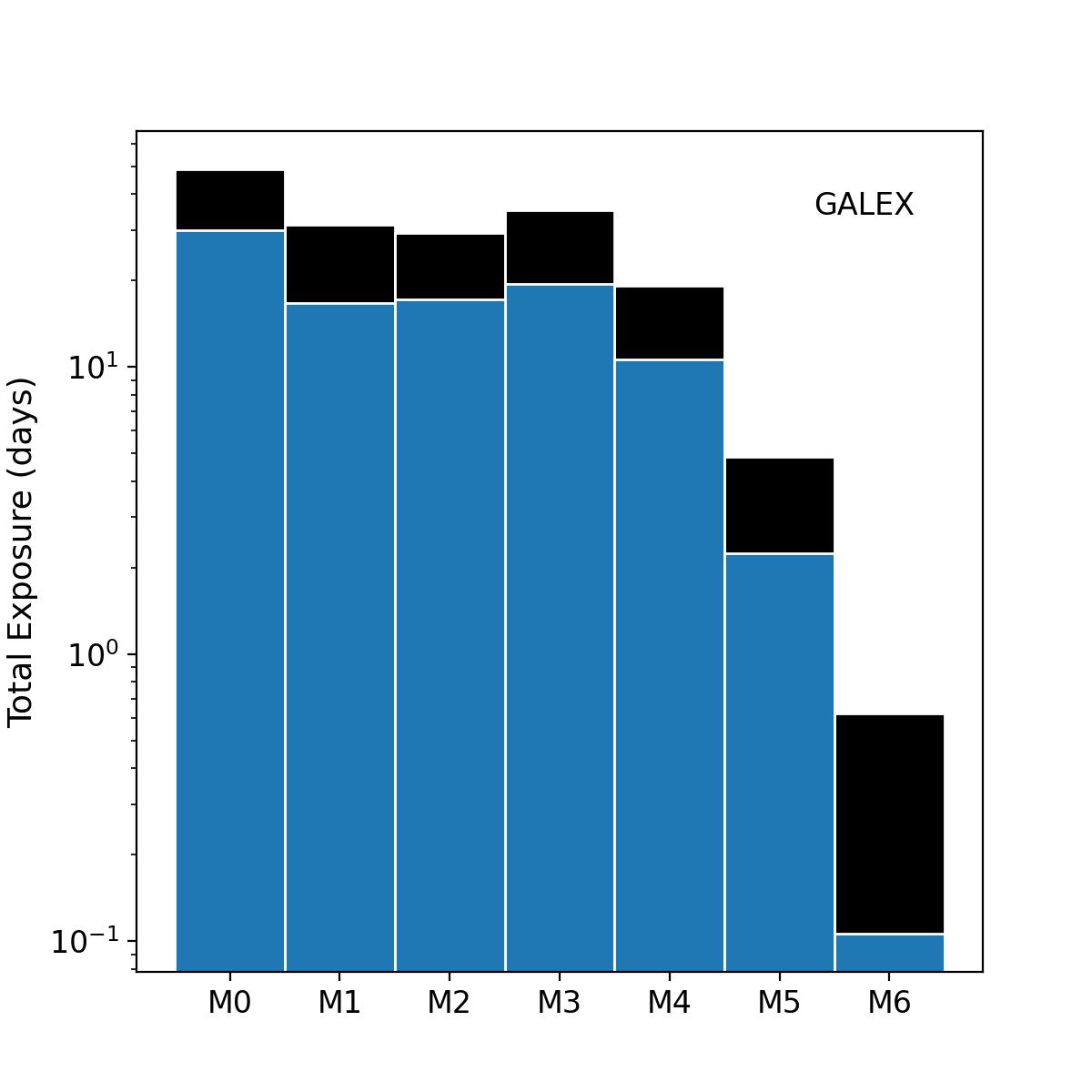}
    }
    \subfloat[]{
    \includegraphics[width=\columnwidth]{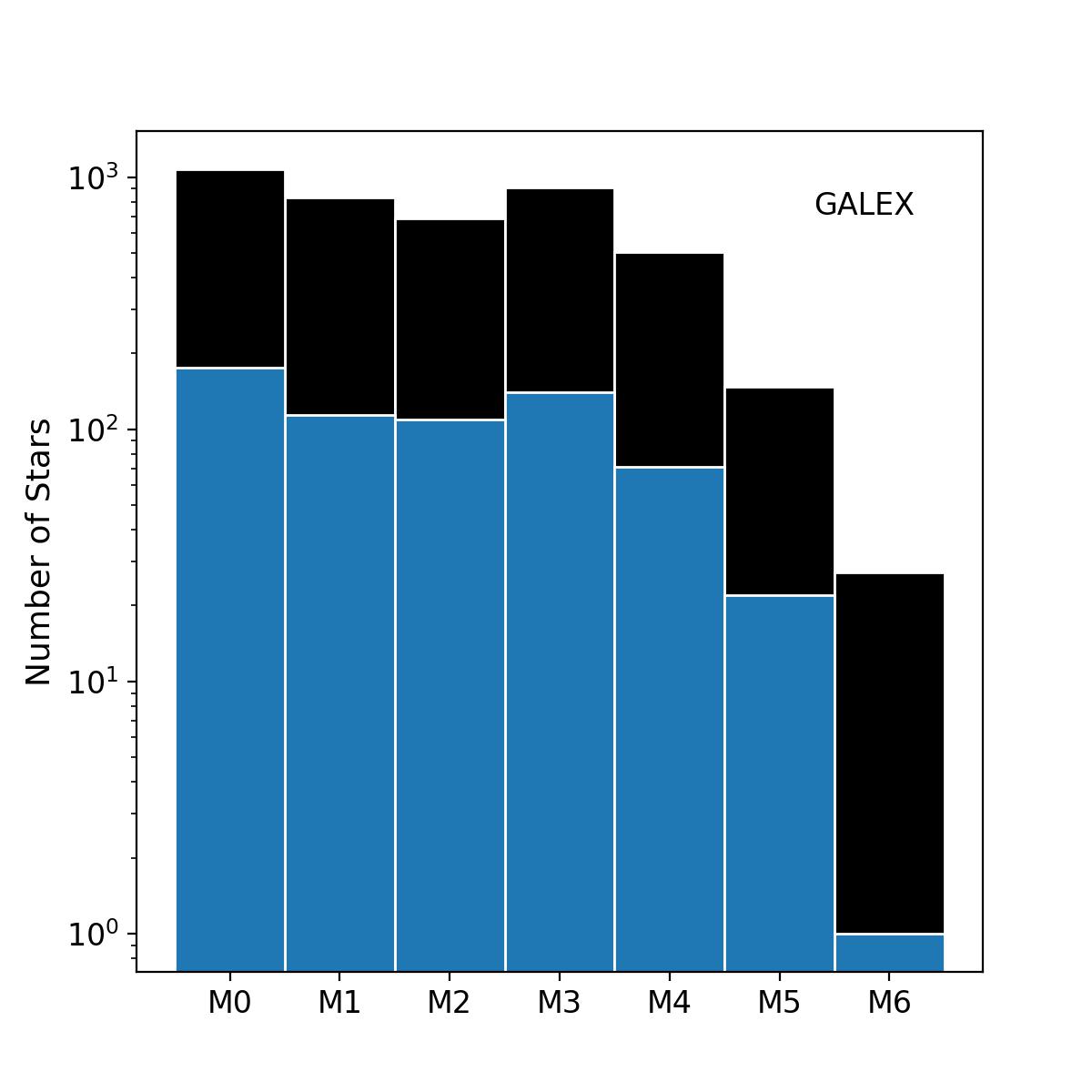}
    }\\
    \subfloat[]{
    \includegraphics[width=\columnwidth]{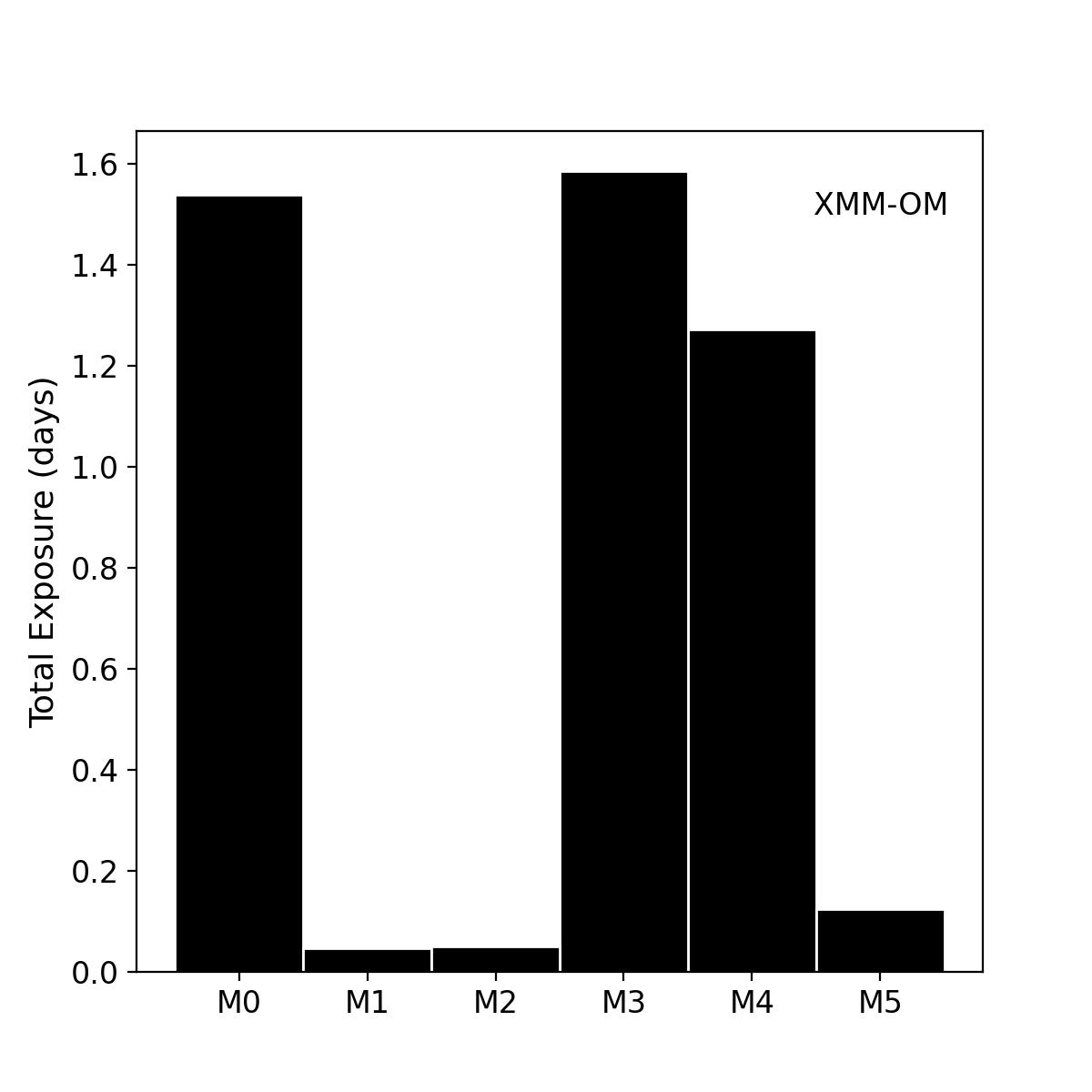}
    }
    \subfloat[]{
    \includegraphics[width=\columnwidth]{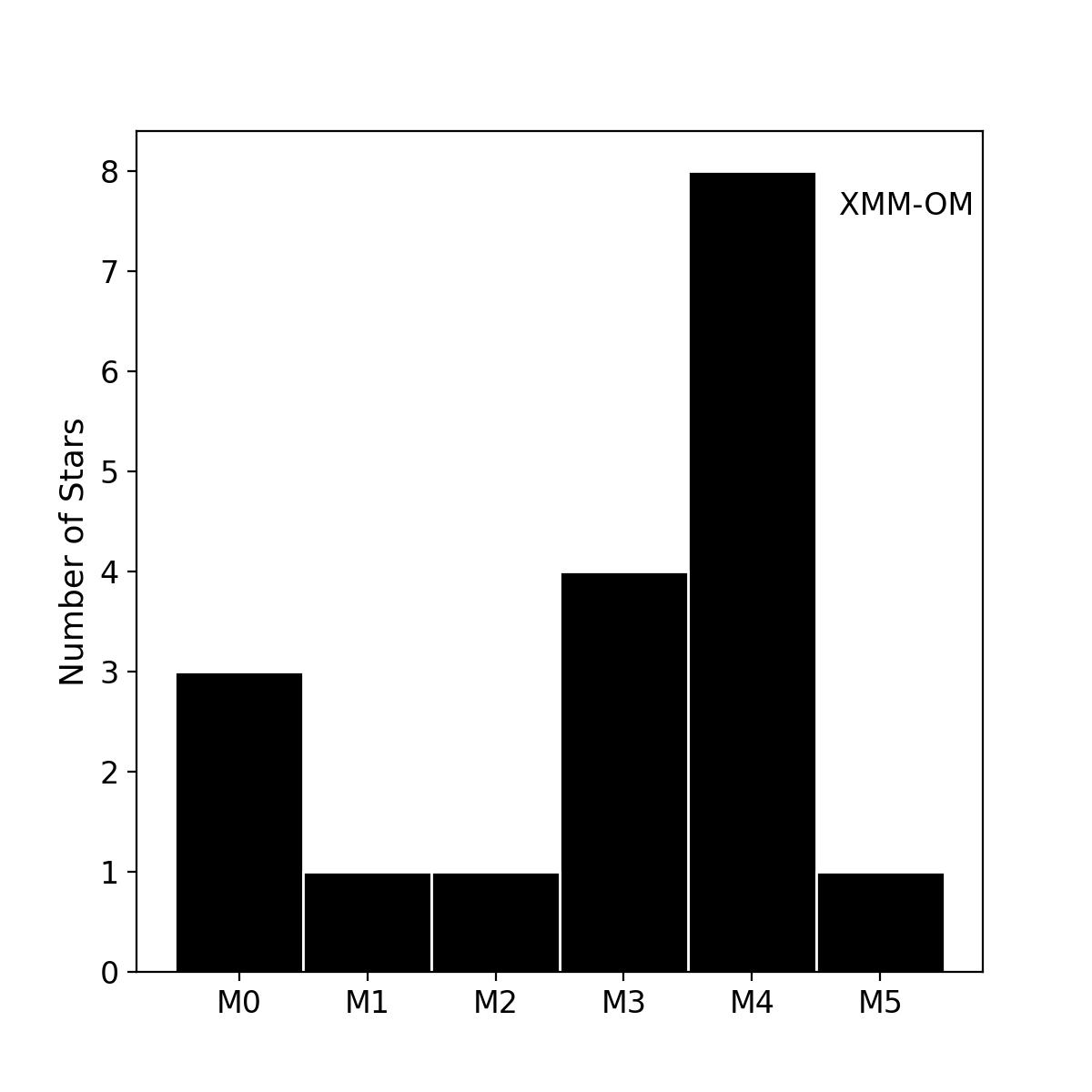}
    }
    \caption{Distribution of our M-dwarf sample as a function of spectral type: (a) Exposure durations and (b) Number of \galex{} sources; (c) Exposure durations and (d) Number of \xmmom{} sources. Black bars show all selected sources on the main sequence as described in section \ref{sec: Dataset}. On the top panels, blue bars show the subgroup of \galex{} sources detected over at least 3 exposures.}
    \label{fig: sources hist}
\end{figure*}

\subsection{\galex{} and gPhoton2 pipeline} \label{sec: galex dataset}

The Galaxy Evolution Explorer \citep{martin_galaxy_2005} is a NASA space-borne telescope that surveyed the sky in the ultraviolet over ten years between 2003 to 2013.
GALEX carried two micro-channel plate detectors with $1.25^\circ$ fields-of-view (FoV), observing in broad ultraviolet bands centered at 1539 \AA{} (Far Ultraviolet; FUV: $1344-1786$ \AA{}) and 2316 \AA{} (Near Ultraviolet; NUV: $1771-2831$ \AA{}) \citep{morrissey_calibration_2007}.
During regular operation the detectors recorded individual photons with a time resolution of $5\,$ms and an angular resolution of $4-6$ arcsec. The time-stamps and positions of every single photon detected are stored in the Mikulski Archive for Space Telescopes\footnote{\url{https://archive.stsci.edu/}} (MAST). \galex{} conducted observations only during the night side of each orbit (an “eclipse”), which could last for up to 1800\,s of the 5900\,s orbit duration. An eclipse could consist of multiple short observations of different positions on the sky or a long observation of a single position. We restrict our dataset to observations longer than 1000s, in an effort to keep processing time to manageable levels without significantly compromising the dataset.

As the original mission pipeline of \galex{} did not release calibrated lightcurves as a standard data products, there exists a vast amount of data unexplored at high time resolutions, which can yield a large dataset of NUV M-dwarf lightcurves. The \gphoton{} pipeline is capable of generating calibrated lightcurves from the \galex{} photonlists stored in MAST, with user-specified time resolutions, which provides the opportunity to exploit this dataset. \gphoton2\footnote{\url{https://github.com/MillionConcepts/gPhoton2}} \citep{st_clair_gphoton2_2022} is a substantial rewrite of the original \gphoton\ library intended to improve deployment flexibility and help support complex, full-catalog surveys on short timescales. It is roughly two orders of magnitude faster than \gphoton\ and features several significant calibration improvements. It is currently available as an early beta release.

The \gphoton2 pipeline works at the level of individual eclipses. It utilises the raw scientific data (\code{raw6}) files from the MAST archive to generate \code{photonlist} files, which record information such as the raw and calibrated detector event positions and time-tags, sky-projected event positions, and a flag that encodes metadata on each detected photon. More details on these files can be found in \cite{million_gphoton_2016}. The \code{photonlist} files can be used to generate sequences of images with user-specified exposure durations. To generate lightcurves, the pipeline performs photometry on each frame in the sequence. We exclude eclipses having operational modes with multiple boresight pointings (`petal-pattern' or AIS modes), as \gphoton2 is not optimised for these modes at the time of writing\footnote{Private correspondence with the \gphoton\ team.}.

The photometry module of \gphoton2 is limited in scope, only able to use circular apertures, and without the ability to identify sources. We thus use the image segmentation package from \noun{photutils}\footnote{\url{https://photutils.readthedocs.io/en/stable/segmentation.html}} to readily distinguish source regions in the images as well as to perform photometry. As a first step, we identify source and sky background regions in the co-added image of the full exposure. We denote source regions as groups of at least 10 connected pixels having flux above 1.5 times the rms background (recommended by \noun{photutils}), with adjoining sources being differentiated via deblending; see \noun{photutils} documentation for detailed information on this process. The minimum of 10 pixels per source matches with \galex{} FWHM of 5.3 arcsec (area of 9.8 pixels with pixel scale of 1.5 arcsec/pixel)\footnote{\url{https://archive.stsci.edu/missions-and-data/galex}}, ensuring no sources are missed.
M-dwarf targets in the image are identified by cross-matching the detected source positions with the M-dwarf list compiled from the TIC within a radius of 3 arcsec, following the procedure used by \cite{bianchi_matched_2020} for construction of the \galex{} - \gaia{} DR2 matched source catalog. These result in a list of 8577 sources, of which 8284 sources fall within the main sequence bounds shown in Figure \ref{fig: HRD}. These are further filtered as per quality cuts described next. 

In each eclipse, we restrict ourselves to targets found in a 1.1 degree diameter FoV instead of the full $1.25^\circ$ FoV, as recommended in \cite{million_gphoton_2016}. In order to avoid contamination, we discard matched sources where the separation between the matched source and the nearest neighbor is less than 3 FWHM of the gaussian PSF fitted by \noun{photutils}. Finally, we discard sources where 10\% or more of the source region pixels were flagged by \gphoton2 in the co-added image. Considering all eclipses longer than 1000\,s, our sample is reduced to 4938 stars after these cuts. Targets that have a neighbouring source in the TIC within 3 arcsec would appear as a single source with \galex{}. These are also discarded to ensure the integrity of our sample, further reducing the sample to 4493 stars. 

\begin{figure}
    \centering
    \includegraphics[width=\columnwidth]{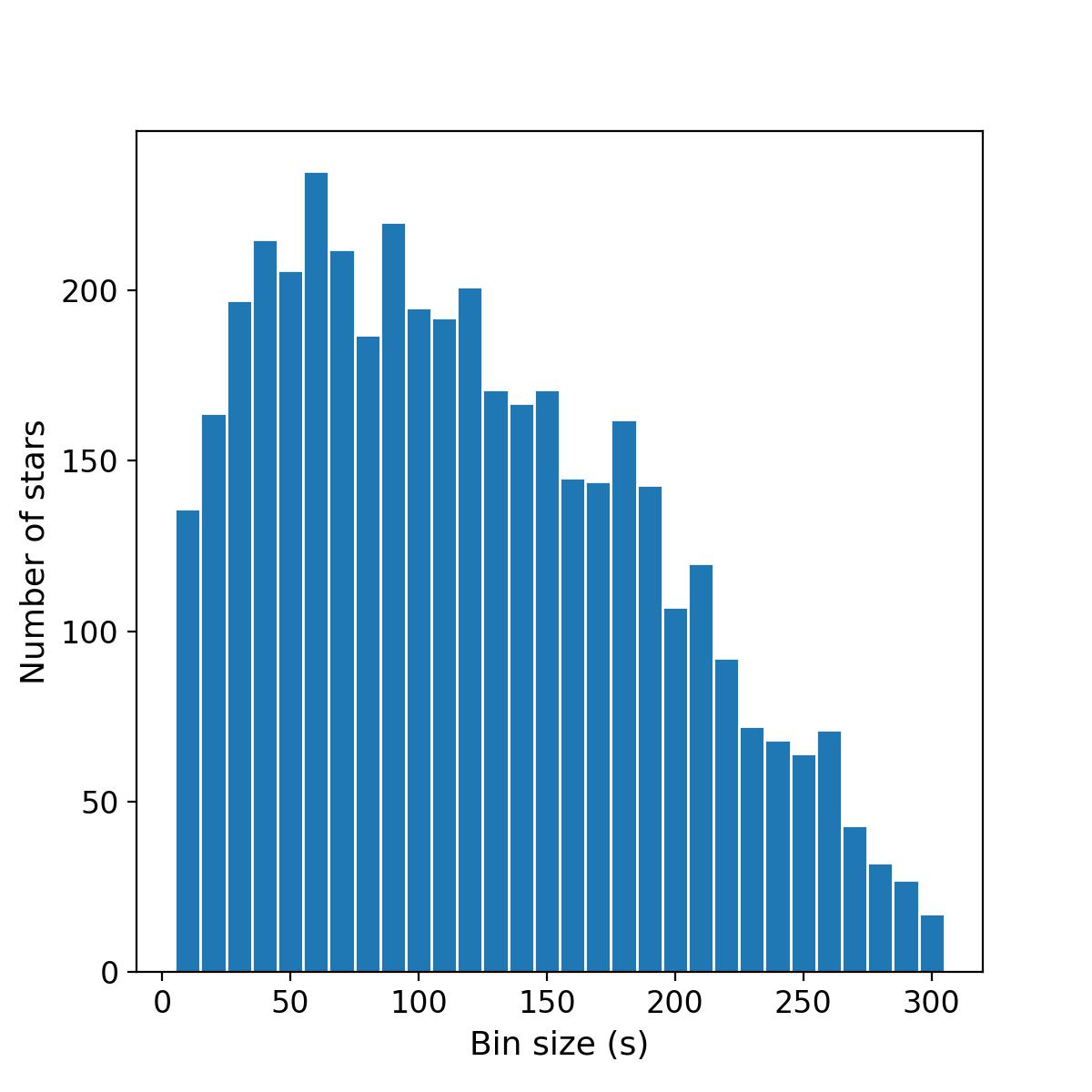}
    \caption{Distribution of bin sizes in the GALEX sample. The shortest detectable flare duration is equal to the bin size as a flare can comprise of a single lightcurve bin (see section \ref{sec: flare id}).}
    \label{fig: bin size hist}
\end{figure}

We generate raw lightcurves for the identified M-dwarf targets by performing photometry on the target regions in each sequence frame. The local background is estimated from a 200 pixel (5 arcmin) box around each target, excluding regions assigned to sources as described previously. The standard deviation of the flux in each time bin is taken as the Poisson errors on raw and background counts, summed in quadrature \citep{million_gphoton_2016}.
Lightcurves are binned such that the nominal Poisson $\text{SNR}~(=\sqrt{\text{S}}$) of a time bin would be 3, where S is the mean count rate multiplied by the bin size. The bin sizes are rounded up to the nearest multiple of 10\,s. The mean count rate for a given source was obtained after masking all suspect and anomalous regions detected upon running the flare identification algorithm with the lightcurves binned in 30\,s durations (see section \ref{sec: flare id} for more details). During this process we discard all sources with a mean count rate less than 0.03 (corresponding bin width greater than 300\,s), or having lightcurves with less than 5 points left after masking suspect and anomalous regions. At this stage, we find 934 sources were detected in only 1 \galex{} exposure, while the relevant fields were observed more than once. Most of these sources had quiescent fluxes near the detection limits, hence having non-detections in other exposures explainable by differing observing conditions. However, 276 of these sources had estimated count rates above 0.1 counts-per-second and may have been detected solely due to flaring activity; these are discarded to prevent contamination of our sample.

Our final GALEX sample consists of 4176 stars. Figure \ref{fig: sources hist} shows the number of stars and observation durations for each spectral type in our sample. Figure \ref{fig: bin size hist} shows the distribution of bin sizes sample. 
446 of these stars had a spectral type classification in SIMBAD \citep{wenger_simbad_2000}, out of which 33 had an `e' classification, indicating enhanced activity in the optical band.

\subsection{\xmm Optical/UV Monitor} \label{sec: xmmom dataset}
\begin{figure}
    \centering
    \includegraphics[width=\columnwidth]{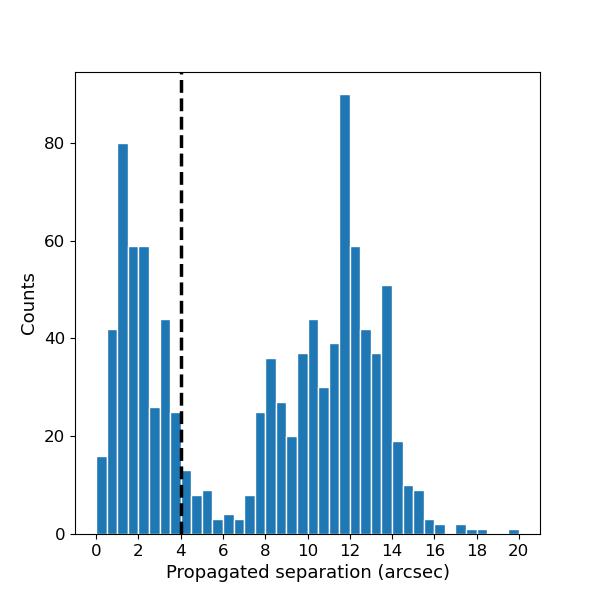}
    \caption{Histogram of closest match separation of \xmmom{} lightcurve positions cross-matched with the TIC sample. The maximum acceptable cross-match is set at 4\,arcsec, shown by the dashed black line.}
    \label{fig:xmmom sep}
\end{figure}

The \xmm Optical/UV Monitor Telescope (XMM-OM) \citep{mason_xmm-newton_2001} is a standalone instrument that provides multi-wavelength observations of XMM targets in the ultraviolet/optical bands simultaneously alongside the X-ray telescope. \xmmom{} can generate high cadence lightcurves in its `fast' mode, which can be used on a single target per exposure. Calibrated and background subtracted lightcurves with 10s bins are generated by the \xmmom{} pipeline and are stored at the XSA. Lightcurves for specified filters can be downloaded in bulk via command line tools. We cross-match all \xmmom{} lightcurves belonging to the three UV filters with the TIC M-dwarf list to identify M-dwarf targets among them. We set the cross-match acceptance threshold to a maximum separation of $4\,$arcsec based on a visual inspection of the cross-match result shown in Figure \ref{fig:xmmom sep}.

\xmmom{} offers three filters in the ultraviolet: UVW1, UVW2 and UVM2. Observations with the UVW2 and UVM2 filters formed a small part of our dataset, and had predominantly low signal-to-noise ratios (SNR<3). We thus restrict our analysis to the observations taken with the UVW1 filter centered at 2931 \AA{} with effective bandpass $2530-3330$ \AA{} \citep{rodrigo_svo_2012, rodrigo_svo_2020}. This sample comprises of 18 stars within the main sequence bounds of Figure \ref{fig: HRD}. Figure \ref{fig: sources hist} shows the number of stars and observation durations for each spectral type in the XMM/OM sample.

The maximum length of a single \xmmom{} fast mode exposure is 4400\,s, with roughly 300\,s gaps between consecutive exposures. Thus each \xmmom{} observation can consist of multiple lightcurves with exposure gaps. We concatenate all lightcurves belonging to a single observation to enable identification of long term trends ($\mathcal{O}(10^3s)$). This helps proper characterisation of flares and a more robust estimation of quiescent flux levels as opposed to analysing each lightcurve individually, see further details below.

\section{Flare Identification and Characterization} \label{sec: flare characterization} 

\subsection{Flare Identification} \label{sec: flare id}
\begin{figure}
    \centering
    \includegraphics[width=5cm]{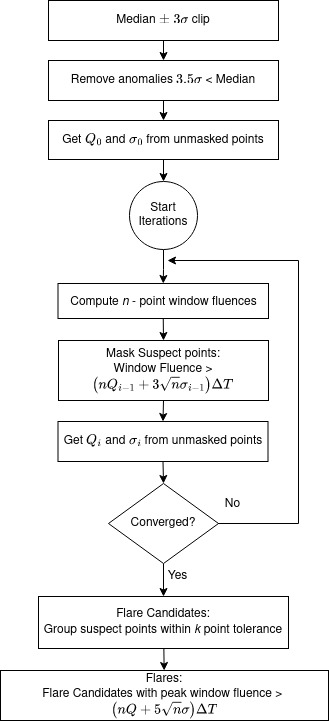}
    \caption{A flowchart detailing the flare identification algorithm used in this work. See text for details.}
    \label{fig:FlowChart}
\end{figure}

Previous studies have shown significant diversity in flare time profiles in the UV \citep{welsh_detection_2007, loyd_muscles_2018, brasseur_short-duration_2019}. Therefore, the identification pipeline must be insensitive to the shape of the flare. Furthermore, data-sets spanning long time periods will inevitably have exposure gaps.
The FLAIIL pipeline\footnote{\url{https://github.com/parkus/flaiil}} developed by \cite{loyd_muscles_2018} specifically targets these challenges. It identifies flares based on the fluence of `runs', which are consecutive points either above or below quiescence. The quiescent flux is modeled using a Gaussian process where fluctuations are described as correlated noise with correlations decaying exponentially with time. For further details, readers can refer to Appendix A of \cite{loyd_muscles_2018}. 
The data contained in our blanket survey of \xmmom{}{} and \galex{}{}, however, has a large number of lightcurves with significantly poorer SNR than the Hubble data utilised by \cite{loyd_muscles_2018}. This, coupled with the short exposure duration of \galex{}{} led us to suspect the `run'-based identification methodology employed by FLAIIL to be not optimal for our dataset. We therefore modify the FLAIIL detection algorithm as described below. A flowchart of the algorithm is shown in Figure \ref{fig:FlowChart}.

We begin by running a sigma clip, where points $>3\sigma$ deviant from the median are masked, where $\sigma$ is the standard deviation of the entire lightcurve around the median. This is done to prevent strong flares from affecting the identification procedure. Points $3.5\sigma$ below (i.e. fainter than) the median are flagged `anomalous' and are permanently masked. These may occur due to instrumental defects.
The unmasked points are used to determine a preliminary estimate of the quiescent flux. $\sigma$ is now updated to be the rms deviation of the unmasked points about the quiescent curve.

Here, we use different strategies to determine the quiescent flux for \galex{} and \xmmom{} lightcurves. A large fraction of the \galex{} sources are observed in only 1-2 exposures. Furthermore, \galex{} has a short single exposure duration (<1800\,s), coupled with much larger gaps between consecutive exposures (>4100\,s) - if present. This makes detrending \galex{} lightcurves unfeasible. Hence the quiescent flux is simply set to be the mean of the unmasked points. On the other hand, \xmmom{} sources have multiple near-continuous exposures, making detrending possible. Hence, \xmmom{} lightcurves have a time-dependent quiescent flux estimate modelled using the Gaussian process module from FLAIIL described previously. We further discuss the implications and workarounds for not detrending \galex{} lightcurves in section \ref{subsec: galex flares}.

Next, for each point in the lightcurve, we compute the fluence in a window of $n$ points around it. We compare this fluence to the quiescent one in the window and flag points with a window fluence $3\bar{\sigma}$ above quiescence as `suspect', where $\bar{\sigma} = \sqrt{n}\sigma$ is the standard deviation of an $n$ point (fluence) sum.
The flagged points are then masked and the quiescent flux and $\sigma$ are recomputed from the unmasked points. When the $n-$point window extends into an exposure gap, the missing points are filled in with the quiescent value. This procedure is then iterated till convergence. Note that in each iteration the previous mask is overwritten.

\begin{figure}[t]
\centering
    \subfloat[]{
      \includegraphics[width=\columnwidth]{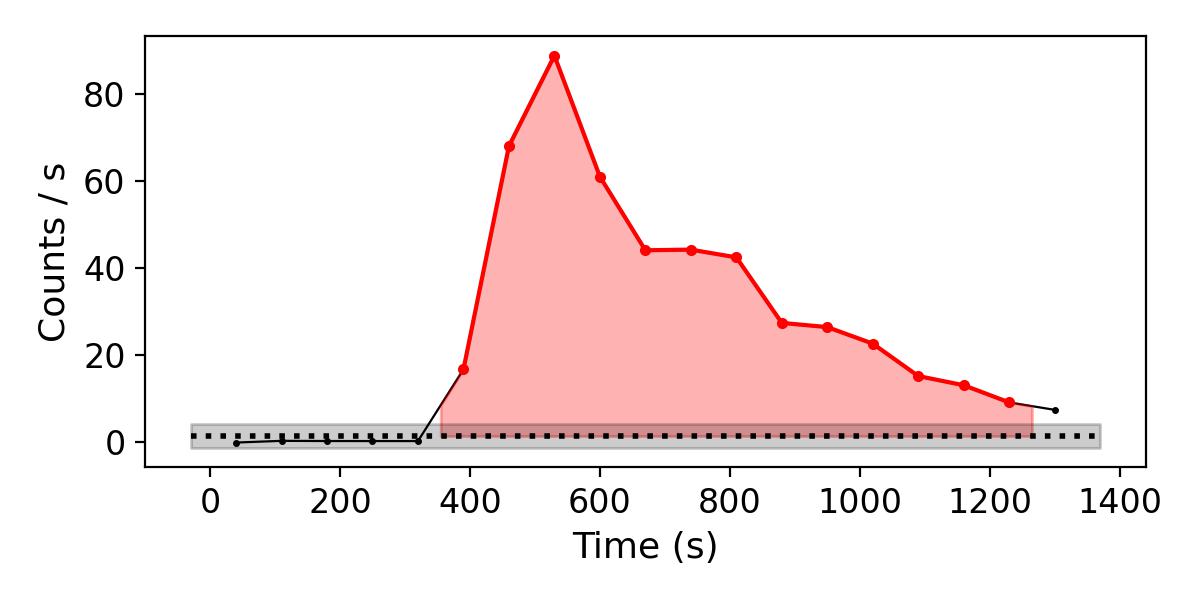}
      \label{subfig:my algo flareid}
    }\\
    \subfloat[]{
      \includegraphics[width=\columnwidth]{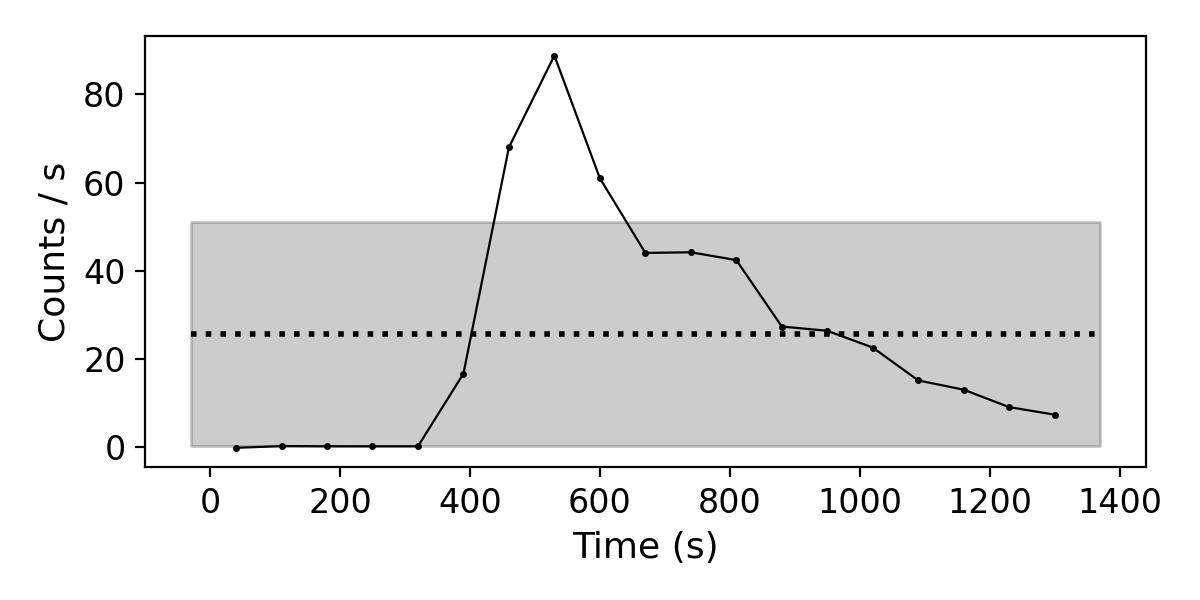}
      \label{subfig:flaiil flareid}
    }
\caption{Comparison of flare detection (shaded red) with our (a) algorithm and (b) FLAIIL for a \galex{} source observed in just one visit. The dotted line denotes the estimated quiescent level, and the gray shaded area is the $\pm 1\sigma$ interval. With FLAIIL, we replace the GP model for the quiescent with a clipped mean. FLAIIL fails to detect the flare and hence gives an erroneously high quiescent estimate.}
\label{fig:flare id comparison}
\end{figure}

We then group consecutive suspect points to create candidate flare ranges. We introduce a $k$ point tolerance for this process wherein ranges within $k$ points of each other are merged. This is useful for efficient detection of the tail of the flare, as well as for detecting small, complex flares in their entirety. We also ensure that no flare candidate extends over exposure gaps or includes anomalous points. Finally, for a range to be confirmed as flaring, we require that at least one n-point window in the range have a fluence $5\bar{\sigma}$ above quiescence. A summary of the flare detections are given in Table \ref{tab: general info}.

Our choice of window size $n$ is based on two criteria: It must be long enough to mitigate short-term noise, yet as short as possible to minimise information loss. By trial and error, we fix $n$ at 3. The tolerance $k$ is set at 3,2,1 and 0 points for bin sizes of 10s, 20s, 30s-40s, >40s respectively.

Due to large variability of flare shapes and the existence of complex flares, we do not make any attempt to identify and separate overlapping flares. We also do not make an effort to reconstruct flares truncated by the start or end of a lightcurve.

Figure \ref{fig:flare id comparison} compares the flare and suspect region identification results using our algorithm and FLAIIL on a \galex{} source detected only in a single visit, as an example. As FLAIIL's quiescent estimation procedure fails when exposure durations are low or when large exposure gaps are present, we replace it with the mean of unflagged points (as described in our algorithm) for this demonstration. FLAIIL requires the endpoints of flare candidates be crossing points, i.e., points where the lightcurve crosses the quiescent curve. This limitation can cause inaccuracies in flare detection and quiescent estimation by FLAIIL in \galex{} data, due to the short visit durations compounded by the inability to construct a quiescent trend. This is exemplified in Figure \ref{fig:flare id comparison}. Furthermore, our methodology offers a more structured framework to determine flare detection thresholds, which is of high relevance for fitting flare frequency distribution as discussed in section \ref{sec: FFD fits}.

\subsubsection{GALEX flares} \label{subsec: galex flares}

\begin{figure}
    \centering
    \includegraphics[width=0.9\columnwidth]{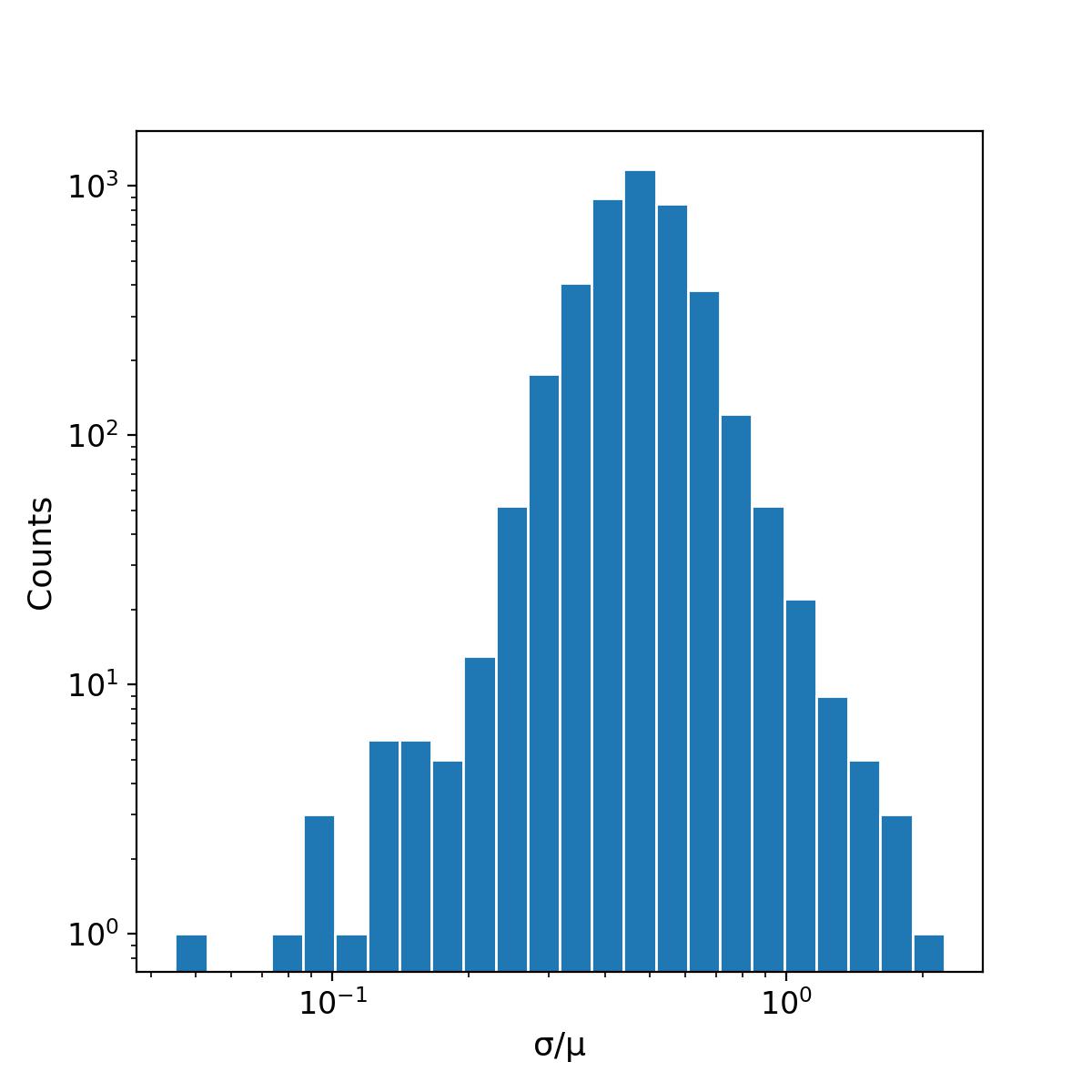}
    \caption{Histogram of the standard deviation to quiescent flux ratio for sources in the \galex{} dataset after concatenation of lightcurves belonging to common sources.}
    \label{fig: nsr hist}
\end{figure}

To accurately estimate the quiescent flux level, \galex{} lightcurves belonging to the same source are concatenated. This is especially helpful in cases where a flare dominates an entire exposure, which makes estimating the quiescent locally challenging. While concatenation can increase the error on the quiescent value - due to long term stellar variability - thereby decreasing our ability to detect smaller flares, we consider the increased certainty of the quiescent a worthy trade-off as it improves our ability to characterize larger flares that are of high interest in the context of planetary atmospheric evolution. Figure \ref{fig: nsr hist} shows the distribution of the standard deviation to quiescent ratios for all sources in the dataset. 
We note here that \galex{} sources detected in only 1 or 2 exposures are susceptible to erroneous estimations of quiescence, due to their lightcurves possibly being dominated by large flares. We elect not to discard these sources, as such large flares are rare, and the sources form a large fraction of our dataset: 43\% in terms of total exposure duration, and 85\% of all sources. We find that 40\% of detected flares occur on these sources, confirming that any bias in flare detection due to errors in quiescence estimation is negligible as the ratio of detected flares to exposure time is near-identical for both sets.

As mentioned in section \ref{sec: galex dataset}, we bin \galex{} lightcurves such that the nominal $\text{SNR}~(=\sqrt{\text{S}}$) of a time bin would be 3, where S is the mean flux multiplied by the bin size. This helps mitigate the effects of noise and enables accurate detection of flares. We obtain an estimate of the mean flux of the source unaffected by flares by an exploratory flare detection run with lightcurves binned in 30\,s bins, recommended by \cite{million_gphoton_2016} as providing a good midpoint in measurement error between the longest and shortest integration. We then re-bin the lightcurves to SNR=3 and run the flare detection algorithm again.

\begin{figure*}
    \centering
    \includegraphics[width=\textwidth]{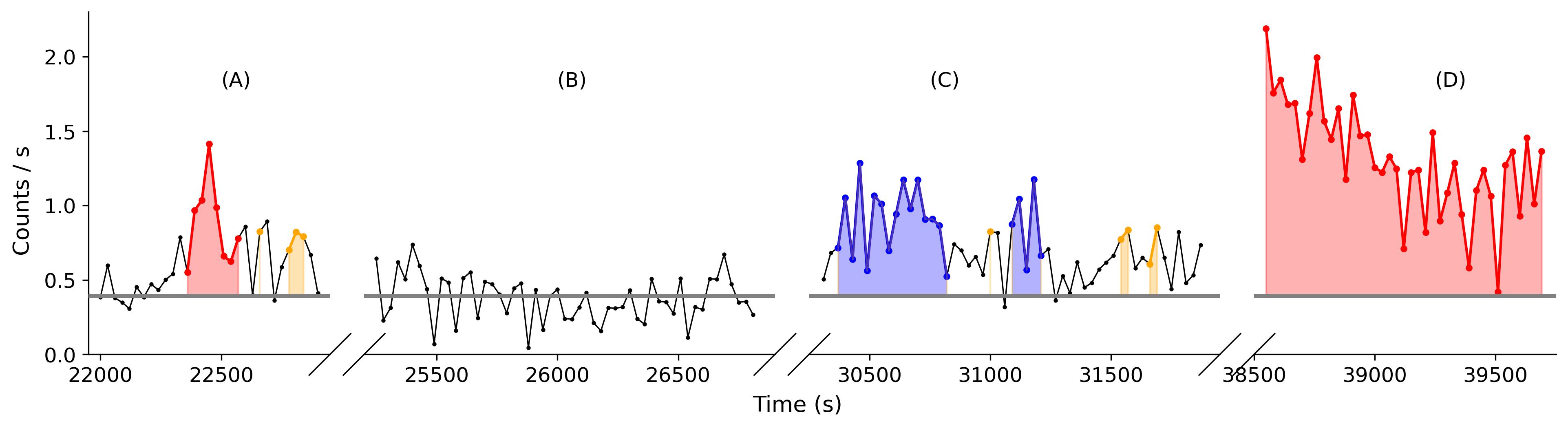}
    \caption{A representative \galex{} lightcurve illustrating the automated flare-detection algorithm and our subsequent manual verification. The red/blue highlights show visually accepted/rejected flares and yellow highlights denote suspect regions that do not qualify as flares (see Figure \ref{fig:FlowChart}). The gray line denotes the estimated quiescent level. 
    (A) shows a small flare with a clear impulse and decay, (B) is a typical 'quiet' lightcurve segment, both blue regions in (C) are rejected for not having impulse or decay phases and being less than 3 times the quiescent and (D) is accepted as a flare for showing a decay phase.}
    \label{fig: galex visual duds}
\end{figure*}

As we cannot model time trends in the underlying quiescent emission in \galex{} lightcurves, the flare detection procedure is susceptible to falsely flagging anomalous regions of increased flux as flares. These may be caused, for example, by rotational modulation or periods of increased magnetic activity. We hence visually inspect all detected flares and exclude those which do not exhibit impulse or decay phases and have an average flux less than 3 times the quiescent level. Our algorithm flagged a total of 593 flares, of which we discard 65. Examples of discarded flares are shown in Figure \ref{fig: galex visual duds}.

Following \cite{brasseur_short-duration_2019} and \cite{jackman_extending_2022}, as well as a private correspondence with the \gphoton\ team, we found that reflection features near the edge of the \galex{} FoV can be falsely detected as flares. These appear as a series of spikes over the entire length of a lightcurve. While the filtering procedure described in section \ref{sec: galex dataset} did not specifically target sources contaminated by these features, they tend to occur near the edge of the \galex{} FoV, and our visual inspection confirmed they get removed.

\begin{figure*}
    \centering
    \includegraphics[width=\textwidth]{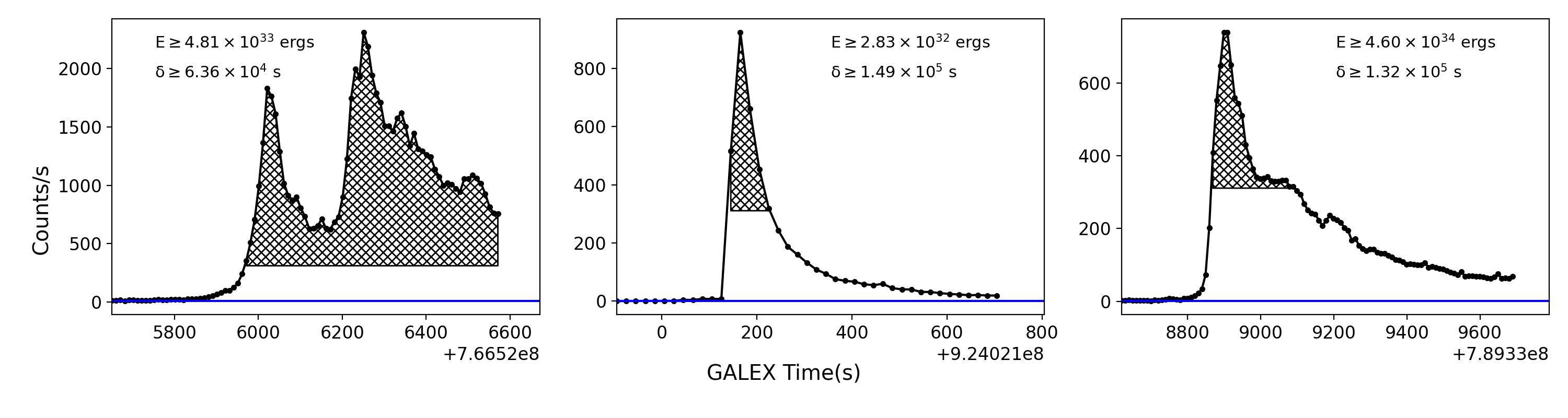}
    \caption{Flares in the \galex{} sample having count rates reaching above 311 cps (shaded region), which is the non-linearity regime of the NUV detector. From left to right, these are observed on the stars TIC 325273691, TIC 359185852, TIC 142215003. The lower limits on flare energy and equivalent duration are shown in the plots.}
    \label{fig: nonlinear flares}
\end{figure*}

Three flares in the \galex{} sample reach fluxes above 311 counts-per-second, and hence go into the non-linearity regime of the \galex{} NUV detector \citep{million_gphoton_2016}. These are shown in Figure \ref{fig: nonlinear flares}. The energy and equivalent duration estimates for these flares are hence lower limits.

The lists of all stars in the sample and all detected and visually verified flares are given in Tables \ref{tab: startable} and \ref{tab: flaretable}.


\subsubsection{XMM-OM flares}
\begin{figure*}
    \centering
    \includegraphics[width=\textwidth]{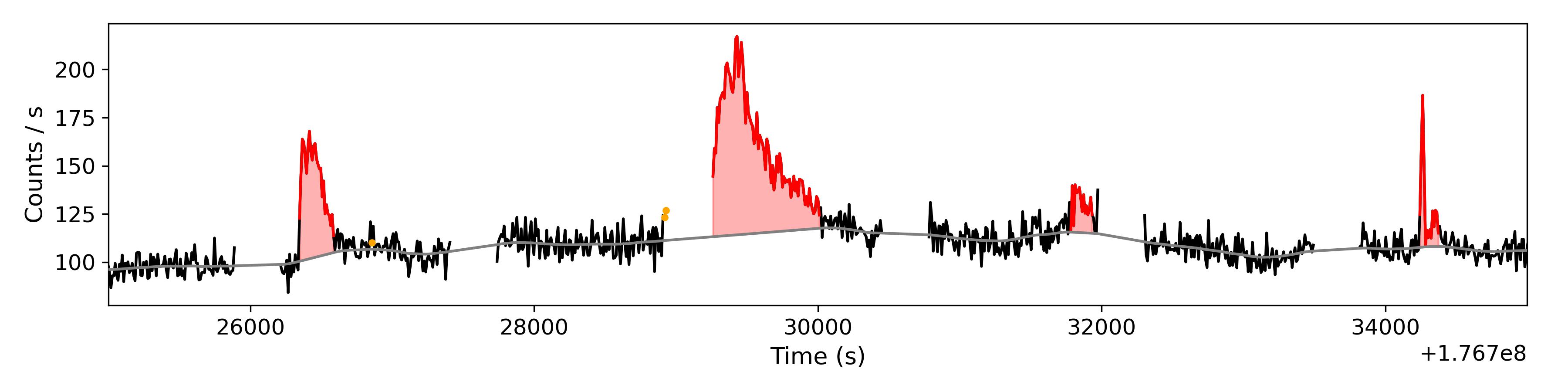}
    \caption{Flares detected on \xmmom{} M0V source TIC 142206123 during observation ID 0148790101. The red regions are flares and suspect points not qualified as flares are plotted in yellow. The gray line shows the quiescent flux estimate modelled with FLAIIL.}   \label{fig: xmmom flare ex}
\end{figure*}

We use pipeline produced \xmmom{} lightcurves unchanged from their 10s binning scheme, as only a minority of the lightcurves have an SNR below 3. Regardless, as seen from Figure 2 and Table \ref{tab: general info}, the total duration of observations and corresponding number of detected flares are too low to merit further analysis.
We also choose not to merge the \xmmom{} and \galex{} flare samples as the additional flare data obtained from the \xmmom{} dataset is not significant enough to warrant tampering with the homogeneity of the \galex{} sample given the difference between their bandpasses and observational methodology.
We have plotted examples of detected flares in Figure \ref{fig: xmmom flare ex}. All detected flares have equivalent durations and energies (section \ref{sec: flare metrics}) consistent with distributions obtained with \galex{}. These results demonstrate the capability of \xmmom{} to monitor M-dwarf flares at high cadences. We list all stars in the sample and all detected flares in Tables \ref{tab: startable xmm} and \ref{tab: flaretable xmm}.

\begin{table}[]
\centering
\begin{tabularx}{\columnwidth}{lcccc}
\hline
SpT & N\textsubscript{stars} & Exposure (days) & N\textsubscript{flares} & N$^\text{flaring}_\text{stars}$\\ \hline
\textbf{GALEX} &  &  &  \\
M0-M2 & 2590 & 109.4 & 171 & 103\\
M3-M6 & 1586 & 59.8 & 357 & 286\\ \hline
\textbf{XMM-OM} &  &  &  \\
M0-M5 & 18 & 4.6 & 20 & 4\\ \hline
\end{tabularx}
\caption{Details of the \galex{} and \xmmom{} flare datasets. Note that the low numbers of flaring stars do not imply that the other stars are inactive, and are rather a consequence of the limited observation duration per star.}
\label{tab: general info}
\end{table}

\subsection{Flare Metrics} \label{sec: flare metrics}
We detail the various metrics we use to characterize flares in the following sections.

\subsubsection{Equivalent Duration}

The equivalent duration, $\delta$, is a measure of flare energy normalized by the quiescent luminosity of the star in a given bandpass. It is defined as  
\begin{equation}
    \delta = \int_{\text{flare}} \frac{(F - F_q)}{F_q} dt,
\end{equation}
where $F$ is measured flux, and $F_q$ is the estimated quiescent flux. As equivalent duration is a measure of flare strength relative to stellar luminosity, it is an ideal quantity for comparing and aggregating flare statistics in a dataset comprised of stars of different ages and activity levels, as well as for comparison between stars of different spectral types; see previous work by \citep{loyd_muscles_2018} and \citep{parke_loyd_hazmat_2018}, which also shows flare frequency distributions as a function of equivalent duration in the FUV to be agnostic to stellar age and activity levels.

\subsubsection{Absolute Energy}

The absolute energy of a flare in the instrument bandwidth is given by
\begin{equation}
    E = 4 \pi d^2 \Delta \lambda_{F} K \int_{\text{flare}} (F - F_q) dt
\end{equation}

where d is the distance to the source, $\Delta \lambda_{F}$ is the effective width of the filter and $K$ is the conversion factor from calibrated counts per second (cps) to energy flux units (erg~s\textsuperscript{-1}~cm\textsuperscript{-2}~\AA{}\textsuperscript{-1}). The effective widths of the NUV and UVW1 filters are 768.31 \AA{} and 795.28 \AA{}, respectively \citep{rodrigo_svo_2012, rodrigo_svo_2020}. $K$ equals $2.06\times 10^{-16}$ for the \galex{} NUV filter\footnote{\url{https://asd.gsfc.nasa.gov/archive/galex/FAQ/counts_background.html}}.  For \xmmom{} we take the value of K for M0V stars given as $1.09 \times 10^{-16}$ \footnote{\url{https://www.cosmos.esa.int/web/xmm-newton/sas-watchout-uvflux}}. The relative errors on the measured distances to the sources in the \galex{} sample peak at 0.084, with an RMS value of 0.007. 
Following previous UV flare studies \citep{loyd_muscles_2018,parke_loyd_hazmat_2018, jackman_extending_2022}, we do not apply a reddening correction to estimated flare energies due to the proximity of our M-dwarf sources ($<200\,$pc).

\subsubsection{Partially Detected Flares} \label{subsec: partial flares}

We categorise flares as incomplete or partially detected if either edge of the flare borders an exposure gap, is $1\sigma$ above the quiescent flux and above 20\% of the peak flare flux. When computing flare frequency distributions we account for incomplete flares via injection/recovery tests as described in section \ref{sec: FFD fits} and Appendix \ref{appendix: injection tests}.

\subsubsection{Flare Length} \label{subsec: flare lengths}

\begin{figure}
    \centering
    \includegraphics[width=\columnwidth]{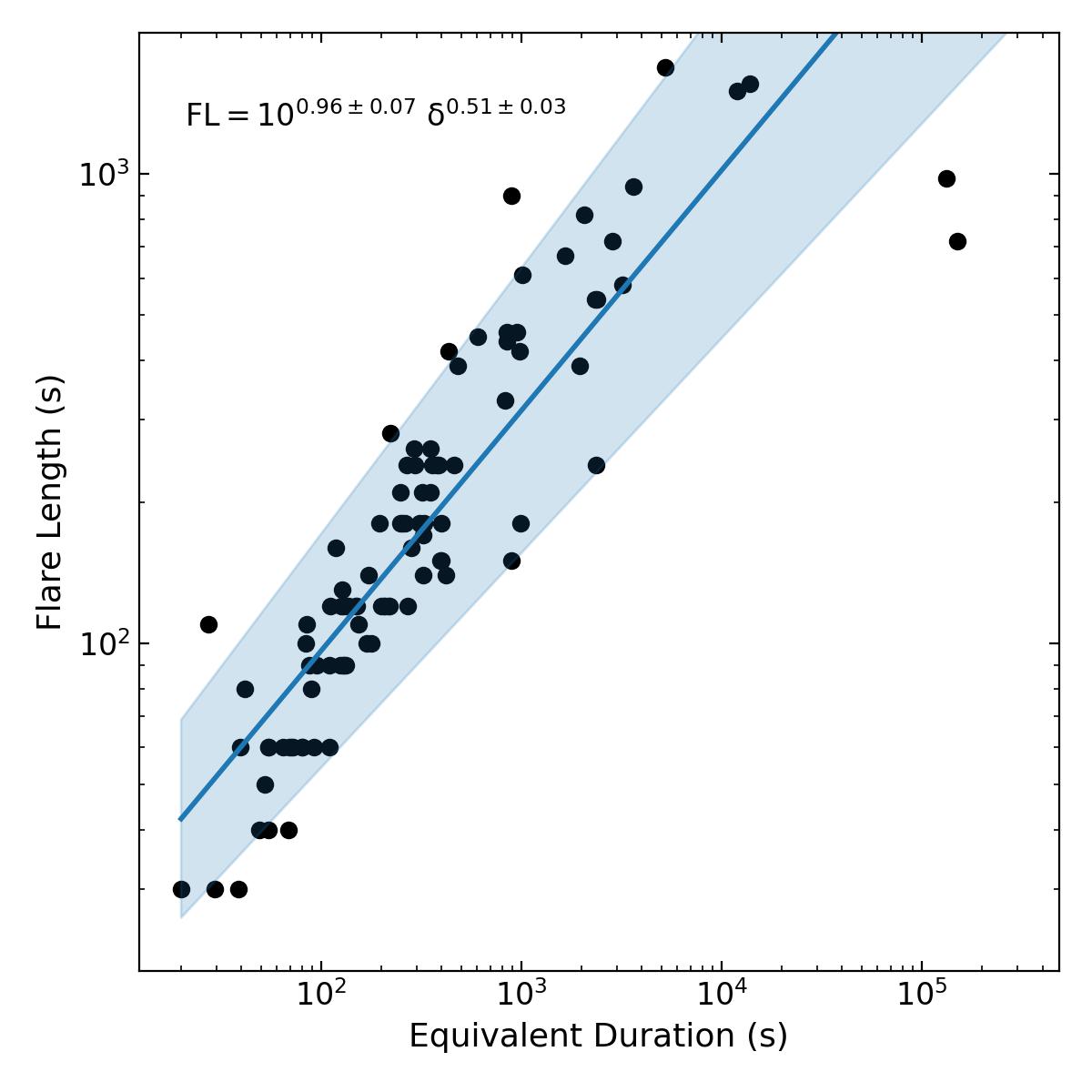}
    \caption{Flare Length as a function of equivalent duration in sources binned at 10\,s to 30\,s. Partially detected flares are excluded from this plot. The solid line shows a power-law fit with $1\sigma$ error region shaded.}
    \label{fig:D vs EqD}
\end{figure}

We define flare length as the time span between the start of the first flaring bin and the end of the last flaring time bin. We find a power-law relationship between flare length and equivalent duration as shown in Figure \ref{fig:D vs EqD}. As flare lengths can be biased to higher values by larger bin widths, we only use flares detected in lightcurves with bin widths between 10\,s and 30\,s to determine this relationship.

\subsubsection{Relative Amplitude}

The relative amplitude of a flare is defined as the average of the flare flux divided by the quiescent flux. It is equivalent to the ratio of equivalent duration to flare length,
\begin{equation}
    \text{Relative Amplitude} ~~ \mathbb{E}\left(\frac{\Delta F}{F_q}\right) = \frac{\delta}{\text{Flare Length}}
    \label{eqn: rel amp - eqd}
\end{equation}




\section{Data Analysis} \label{sec: Data Analysis}

Our analysis of the \galex{} NUV dataset includes two components. The first is a traditional frequency distribution analysis of visually validated flares. The second component is a long baseline activity study which does not depend on detection of individual flare events, but rather looks at the effects of flares and increased activity as a whole. The second component features lightcurves binned in 300s bins, which is the proposed single exposure duration of ULTRASAT \citep[see discussion in section \ref{sec: ultrasat};]{ben-ami_scientific_2022, shvartzvald_ultrasat_2023}.

\subsection{NUV Flare Frequency Distributions} \label{sec: FFD fits}

We analyse the flare data obtained in the previous section to estimate rates of flare activity across M-dwarf spectral types, both as a function of flare energy and equivalent duration. The standard tool used to estimate flare activity is to compute cumulative flare equivalent duration - frequency and energy - frequency distributions (FFDs), which have been shown to follow a power law across a wide range of wavelengths (NIR to UV; \citealt{loyd_muscles_2018, murray_study_2022, stelzer_flares_2022}). We thus have the relations
\begin{equation}
    d\nu(\delta) = -k\delta^{-\beta} d\delta, \qquad  k>0,\beta>1
    \label{FFD eqn}
\end{equation}
and
\begin{equation}
    d\nu(E) = -kE^{-\beta} dE, \qquad  k>0,\beta>1
    \label{FFD eqn Energy}
\end{equation}
where $\nu$ is the occurrence rate of flares with equivalent duration greater than $\delta$, and $k$ and $\beta$ are parameters of the distribution. We employ an MCMC sampler to obtain fits for $k$ and $\beta$. The fit procedure works directly from the discrete flare events (i.e., does not require binning of the observed FFD) and accounts for the varying lower detection limits when events from multiple sources and time periods are aggregated. Although fits to $k$ and $\beta$ are correlated, MCMC sampling is by its nature robust against such correlations \citep{sharma_markov_2017}.

We set lower detection limits on flare fluence (and consequently equivalent duration) for each individual source, arising as a direct consequence of the detection algorithm we employed. As each flare detection needs to have at least a single $n$-point window which is $5\bar{\sigma}$ above the quiescent (section \ref{sec: flare id}), that defines our detection threshold. That is, each lightcurve has a flare detection threshold fluence
\begin{equation}
    F_{low} = 5\sqrt{n}\sigma \Delta t,
\end{equation}
and corresponding threshold equivalent duration
\begin{equation}
    \delta_{low} = F_{low}/\mu, 
\end{equation}
where $\Delta t$ is the bin width, $\mu$ is the quiescent flux level, and $\sigma$ is the rms deviation of the lightcurve. We support this claim with flare injection/recovery tests conducted on a randomly selected subset of lightcurves, described in Appendix \ref{appendix: injection tests}.

We follow the MCMC fitting procedure detailed in appendix B of \cite{loyd_muscles_2018}, which we briefly explain next. We use the Python package \code{ffd} \footnote{\url{http://www.github.com/parkus/ffd}} which utilises this procedure. In the following explanation, we state only the equations for the equivalent duration FFD for brevity as the energy FFD obeys the same relations.  The cumulative distribution function $\nu$ is obtained by integrating Eq. \ref{FFD eqn},
\begin{equation}
    \nu = C\,\delta^{-\alpha}.
    \label{FFD intg}
\end{equation}
\noindent
where $C = k/(\beta-1)$ and $\alpha = \beta -1$. We sample the joint likelihood of the flare rate constant, $C$, and the power law index, $\alpha$. The probability of $n$ flares occurring, $p(n)$, is assumed to be independent of the probability distribution of the flares' equivalent duration, $p(\boldsymbol{\delta})$. Thus, for a single observation, the net likelihood of the data is
\begin{equation}
    p(n,\boldsymbol{\delta}; C,\alpha) = p(n)p(\boldsymbol{\delta}),
\end{equation}
where $\boldsymbol{\delta}$ is the set of equivalent durations of the $n$ flares.

We assume that flare events are always independent, which could be inaccurate yet describes event rates well \citep{wheatland_origin_2000}. Thus, $p(n)$ is given by a Poisson distribution, with the number of expected flares in an observation $N$ determined by Eq. \ref{FFD intg}, using the duration of the observation $\Delta T$ and the lower detection limit $\delta_{low}$,
\begin{equation}
    N = C\Delta T \delta_{low}^{-\alpha}.
    \label{eqn: expected flare numbers}
\end{equation}

\noindent
The likelihood of the observed flare equivalent durations $p({\boldsymbol{\delta}})$ is obtained from the power-law part of Eq. \ref{FFD intg}. After appropriate normalization, this is
\begin{equation}
    p(\boldsymbol{\delta}) = \prod_{i} \frac{\alpha\,\delta^{-\alpha-1}}{\delta_{low}^{-\alpha}}
    \label{eqn: flare delta distribution}
\end{equation}
where $i$ indexes individual flares in the set ${\boldsymbol{\delta}}$. When no events are detected, we have
\begin{equation}
    p(n,\boldsymbol{\delta}; C,\alpha) = p(0,\varnothing;C,\alpha) = \text{Poisson}(0).
\end{equation}

Our datasets includes a multitude of lightcurves with varying durations and detection limits. These observations are independent, and hence the final likelihood of all data used by the MCMC sampler is taken as the product of the data likelihoods for each separate observation.

As \galex{} continuous exposures are shorter than 1800\,s, it is possible that fits to the observed FFD may be biased due to the partial detection of flares, especially high equivalent duration flares (see sections \ref{subsec: partial flares} and \ref{subsec: flare lengths}). This could occur in two ways, with the number of low equivalent duration flares being overestimated and the number of high equivalent duration flares being underestimated. As the boundary between these two effects varies between different sources and different flare profiles, we conduct aggregated injection/recovery tests to estimate correction factors for $\alpha$ and $C$.

\begin{figure*}
\centering
\subfloat{
    \includegraphics[width=\columnwidth]{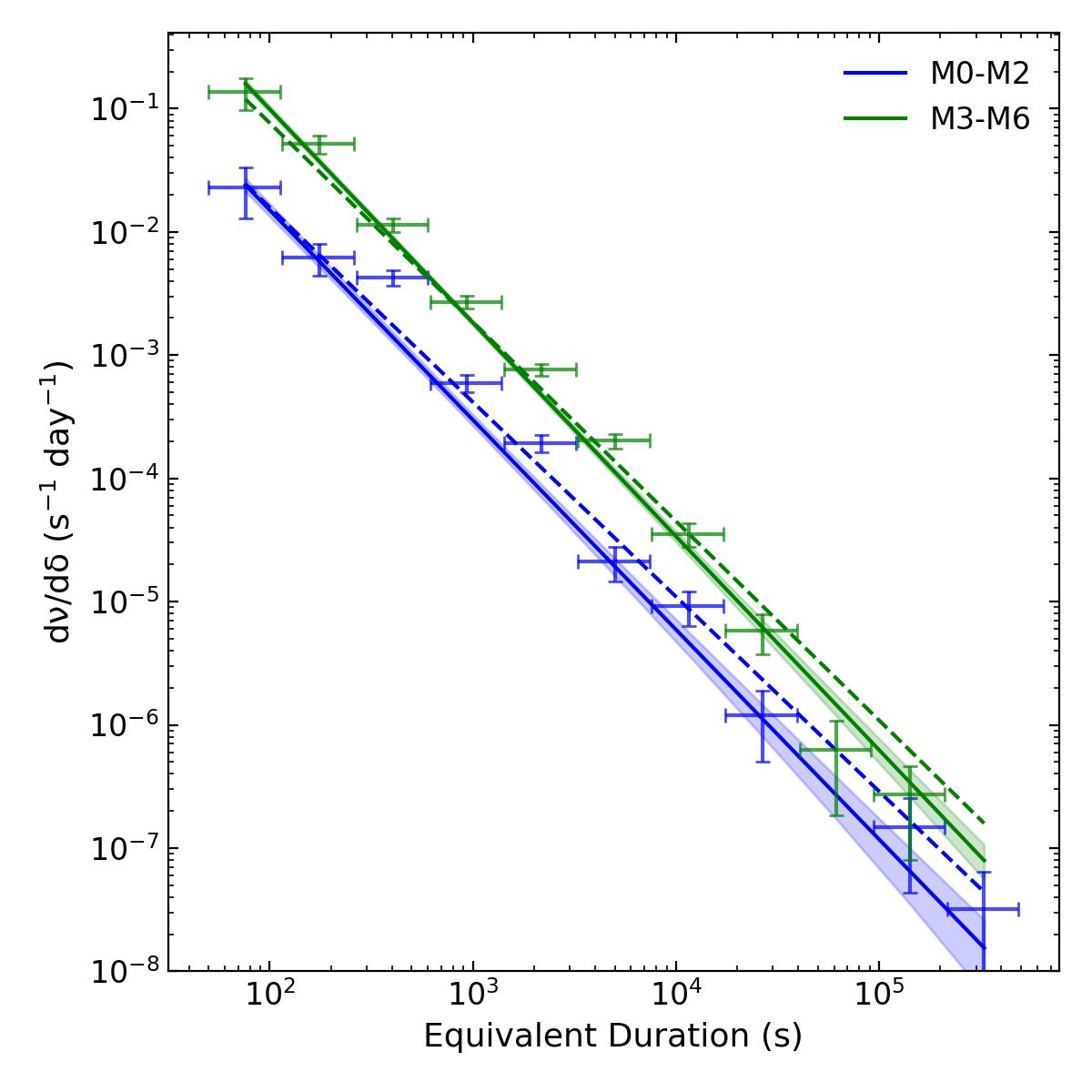}
    \label{subfig: grouped ffd delta}
      }
\subfloat{
    \includegraphics[width=\columnwidth]{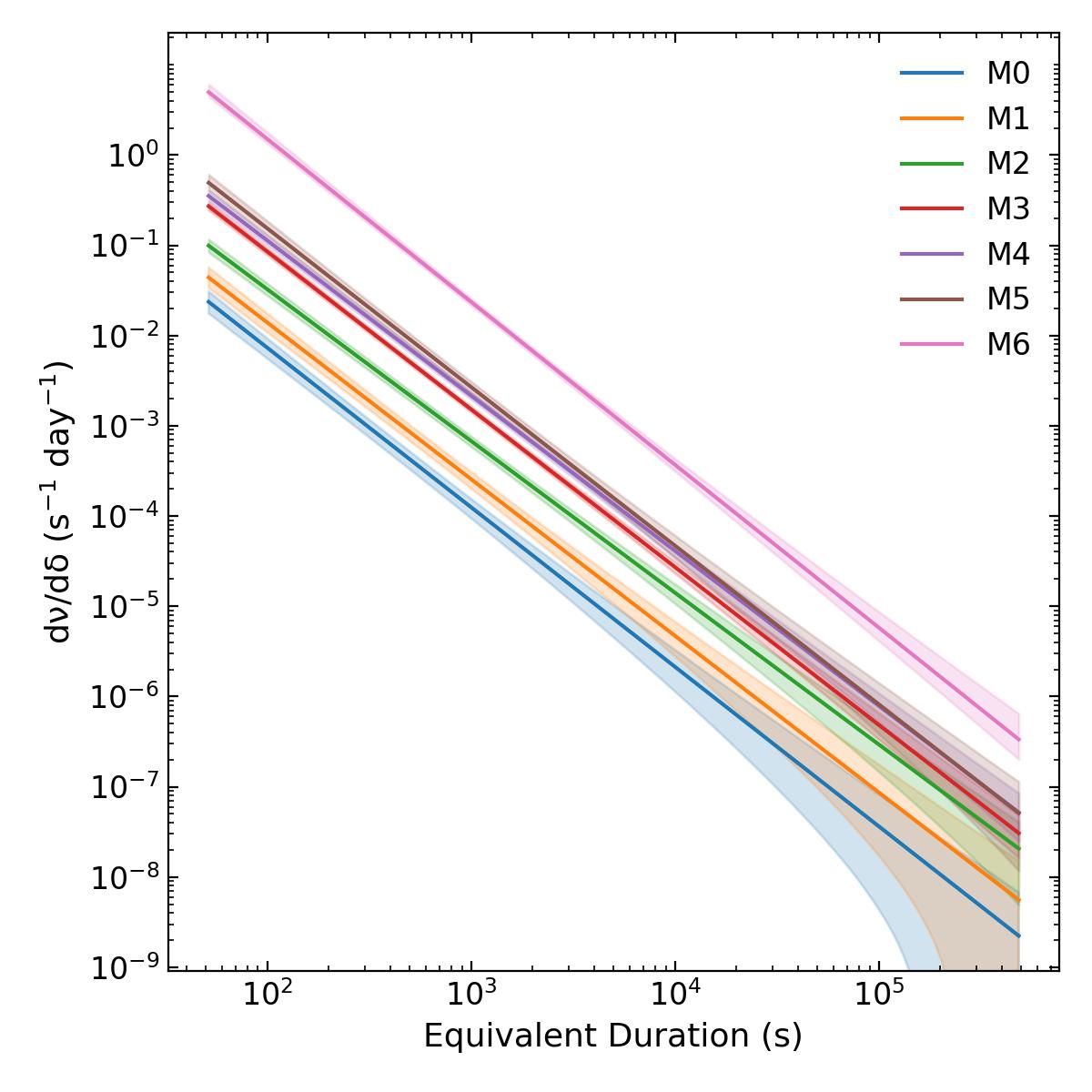}
    \label{subfig: indiv ffd delta}
}
\caption{(a) GALEX observed flare frequency distribution in equivalent duration with $1\sigma$ error bars. The y-axis denotes frequency density $d\nu/d\delta$ as defined in Eq. \ref{FFD eqn}, with the dashed line being the nominal power-law fit. The solid line is the corrected fit as described in section \ref{sec: FFD fits}. (b) Corrected GALEX equivalent duration FFD fits for individual M spectral types. Observed FFDs for individual spectral types are plotted in Figure \ref{fig: eqd indiv fits}. Fit parameters are given in Table \ref{tab: FFD fits}.}
\label{fig: Eq dur FFD}
\end{figure*}

\begin{figure*}
\centering
\subfloat{
    \includegraphics[width=\columnwidth]{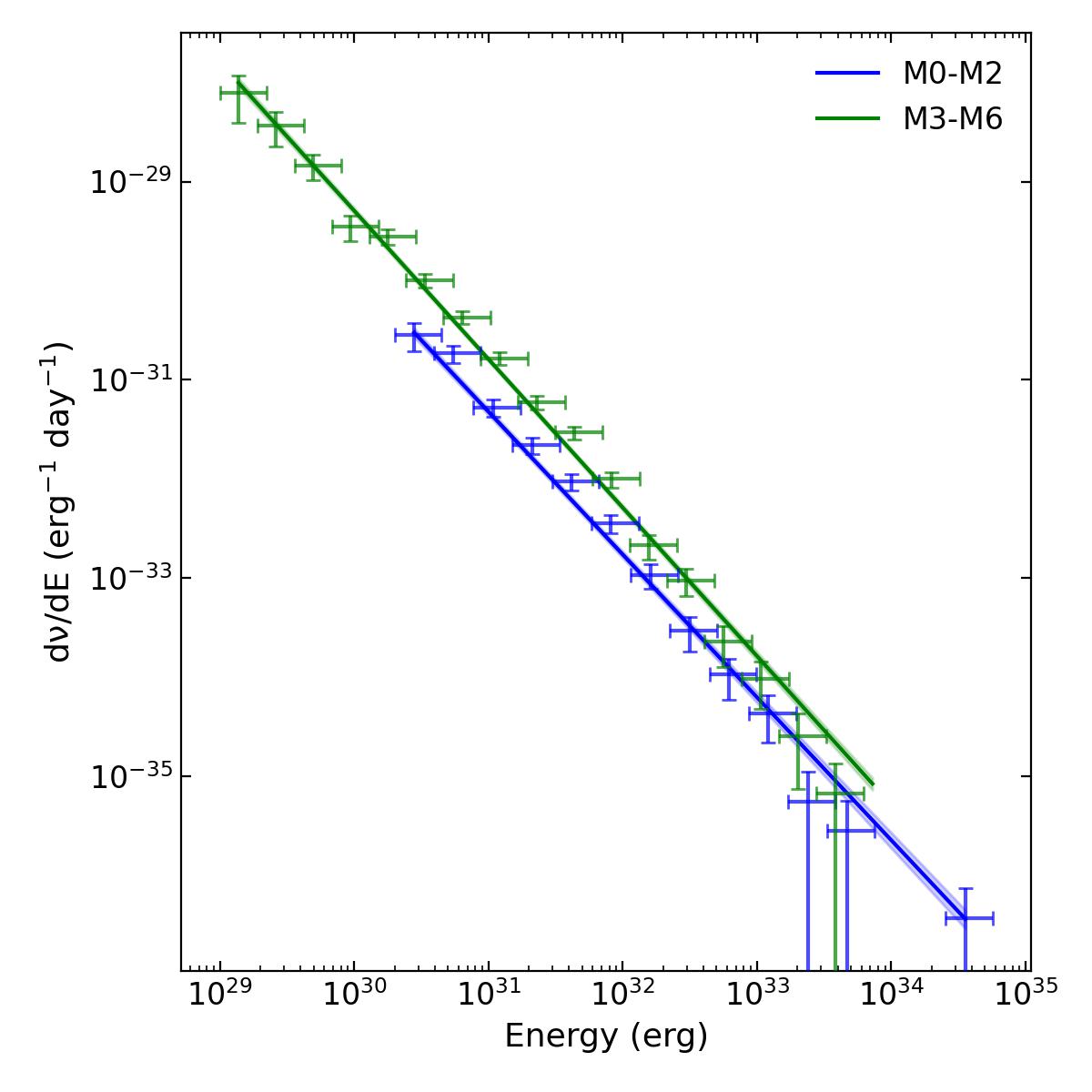}
    \label{subfig: grouped ffd energy}
      }
\subfloat{
    \includegraphics[width=\columnwidth]{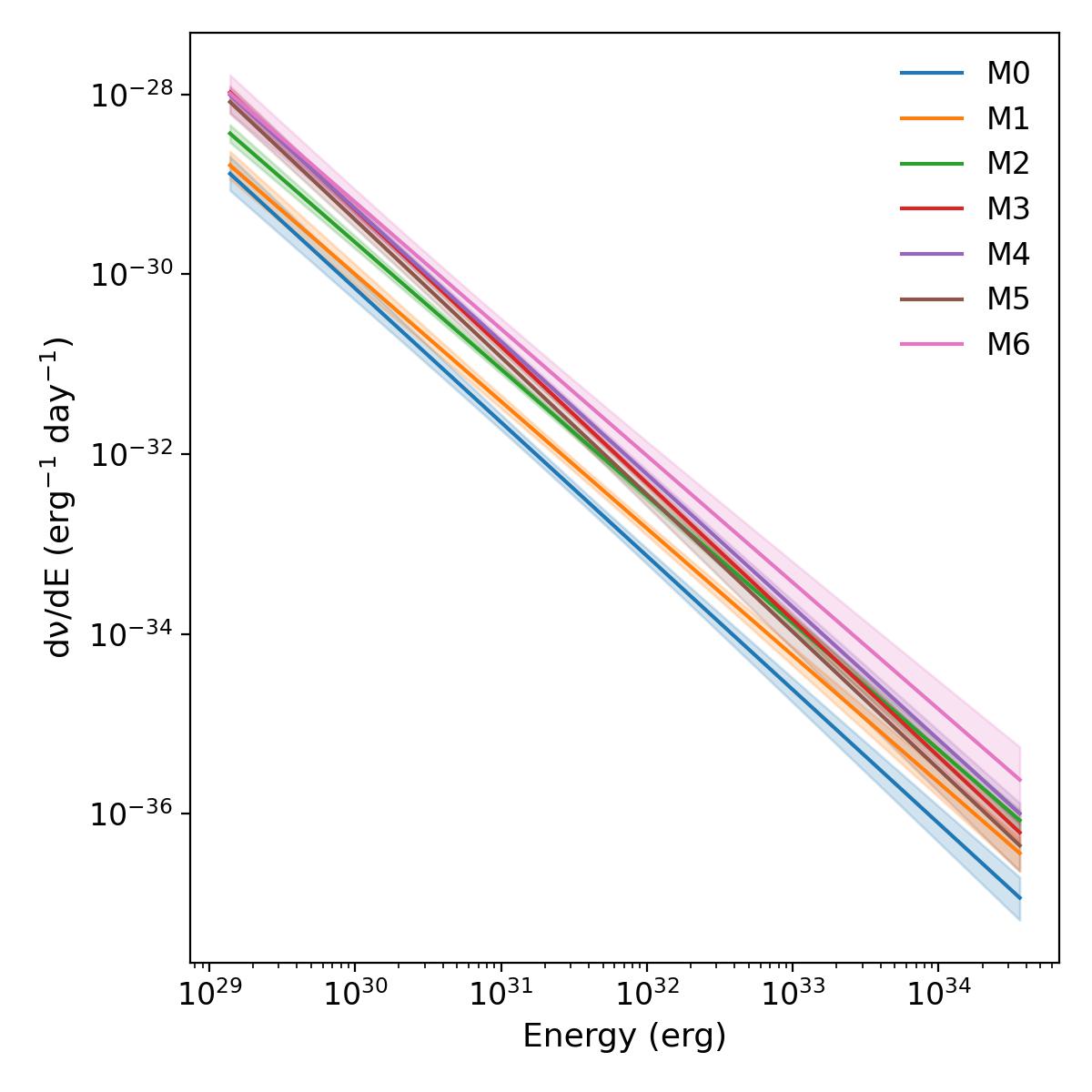}
    \label{subfig: indiv ffd energy}
}
\caption{(a) GALEX flare frequency distribution in energy with $1\sigma$ error bars. The solid line denotes the power-law fit. The y-axis denotes frequency density $dE/d\delta$ as defined in Eq. \ref{FFD eqn Energy}. (b) Energy FFD fits for individual M spectral types. Observed FFDs for individual spectral types are plotted in Figure \ref{fig: energy indiv fits}. Fit parameters are given in Table \ref{tab: FFD fits}.}
\label{fig: Energy FFD}
\end{figure*}

\begin{figure}
\centering
\includegraphics[width=\columnwidth]{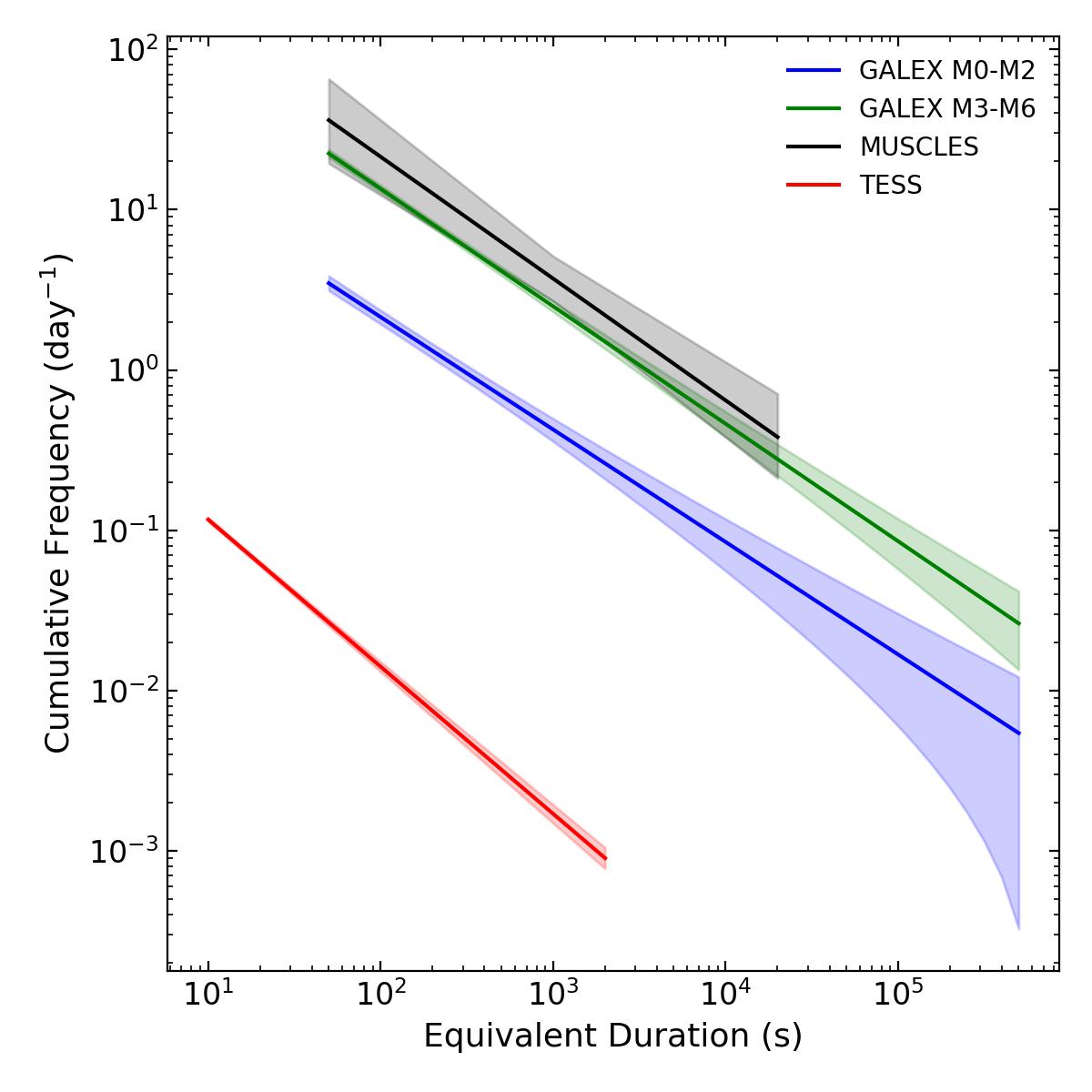}
\caption{Comparison of cumulative flare frequency distributions in equivalent durations. MUSCLES FFD is in the FUV \citep{loyd_muscles_2018}. TESS FFD is computed from Table 2 of \cite{stelzer_flares_2022}. }
\label{fig: compare delta FFDs}
\end{figure}

\begin{figure}
    \centering
    \includegraphics[width=\columnwidth]{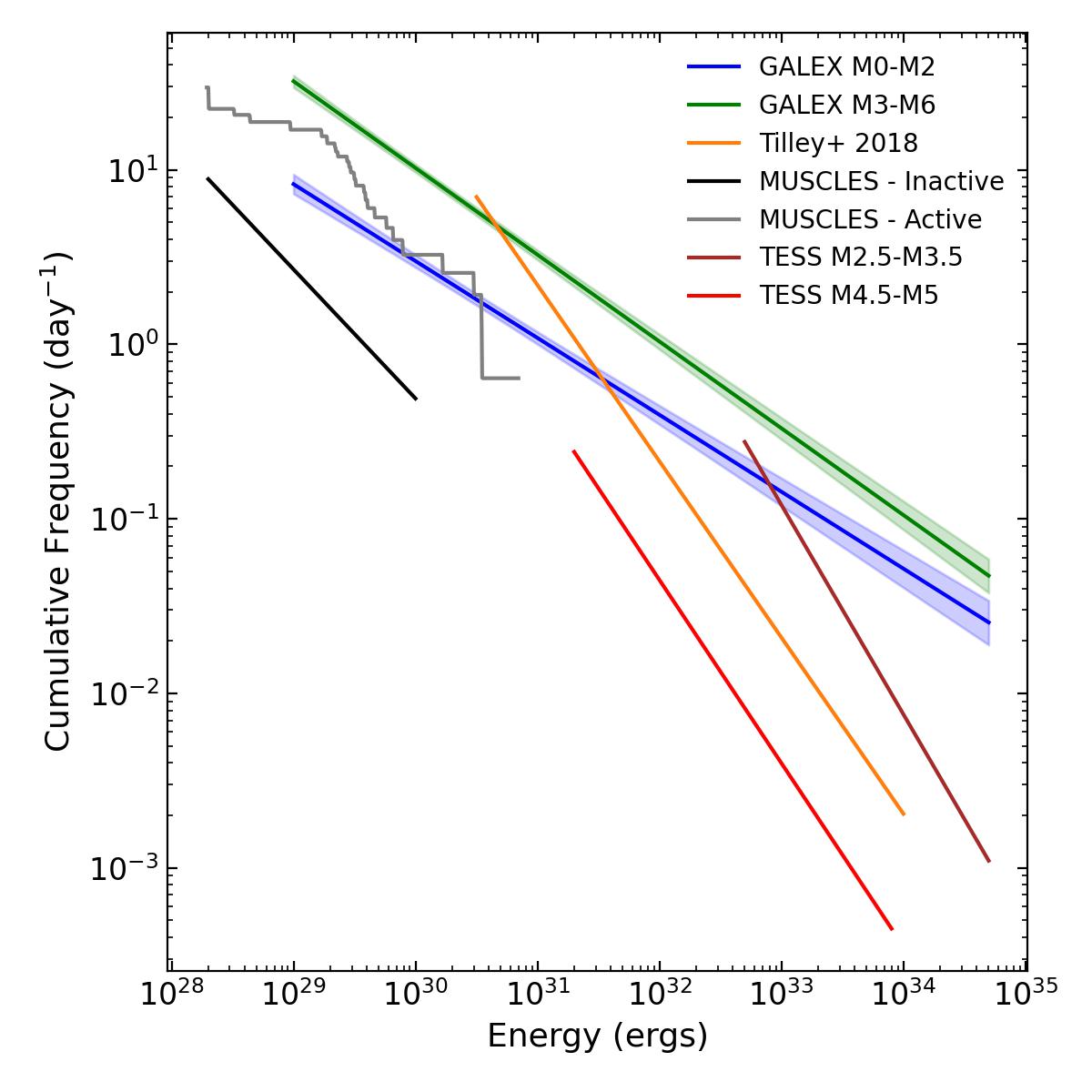}
    \caption{Comparison of cumulative flare frequency distributions in energy. The MUSCLES FFDs are in the FUV \citep{loyd_muscles_2018}. TESS FFDs are obtained from \cite{stelzer_flares_2022}. The \cite{tilley_modeling_2019} FFD is a simulation in the U-band, extrapolated from Kepler observations by \cite{hawley_kepler_2014}.}
    \label{fig: energy ffd comparison}
\end{figure}

We obtain FFDs and power-law fits for each individual M-dwarf sub-type, from M0 to M6. We also do the same for two broader categories: early M-dwarfs (M0-M2) and mid M-dwarfs (M3-M6) to obtain improved confidence intervals. This division is grounded on M-dwarfs turning fully convective at M3-M4 \citep{chabrier_structure_1997, baraffe_closer_2018}.The nominal and corrected fits for the FFDs are shown in Figures \ref{fig: Eq dur FFD} and \ref{fig: Energy FFD} and in Table \ref{tab: FFD fits}. We do not attempt to correct the energy FFD fits, as that has an additional variable in the quiescent luminosity of the source itself. As seen from Figure \ref{fig: Eq dur FFD}, the magnitude of the shift in the equivalent duration FFD after correction is relatively small, hence we do not expect the energy FFD to significantly deviate from the true distribution. We also note that while three flares in the sample reach the non-linearity regime of the NUV detector (Figure \ref{fig: nonlinear flares}), they cannot shift our FFDs beyond the error bars due to the small number of affected flares. We confirmed this by conducting fits after artificially increasing the fluence of these flares up to a factor of 10. We show comparisons to cumulative FFDs obtained from HST \citep{loyd_muscles_2018} and TESS \citep{stelzer_flares_2022} data in Figures \ref{fig: compare delta FFDs} and \ref{fig: energy ffd comparison}. We discuss these results in section \ref{sec: Discussion}.

\begin{table}[]
\centering
\begin{tabular}{ccccc}
\hline
SpT & \multicolumn{2}{c}{Power-Law Index} & \multicolumn{2}{c}{Log Constant} \\
 & \multicolumn{2}{c}{$\alpha$} & \multicolumn{2}{c}{log\,C} \\ \hline
 & \multicolumn{4}{c}{Equivalent Duration} \\
 & Nominal & Corrected & Nominal & Corrected \\ \hline
M0 & $0.66 \pm 0.09$ & $0.77 \pm 0.07$ & $1.38 \pm 0.20$ & $1.52 \pm 0.21$ \\
M1 & $0.62 \pm 0.08$ & $0.74 \pm 0.07$ & $1.61 \pm 0.19$ & $1.75 \pm 0.20$ \\
M2 & $0.56 \pm 0.05$ & $0.68 \pm 0.04$ & $1.88 \pm 0.12$ & $2.04 \pm 0.14$ \\
M3 & $0.63 \pm 0.04$ & $0.75 \pm 0.03$ & $2.29 \pm 0.10$ & $2.55 \pm 0.13$ \\
M4 & $0.60 \pm 0.04$ & $0.72 \pm 0.04$ & $2.35 \pm 0.12$ & $2.63 \pm 0.16$ \\
M5 & $0.65 \pm 0.09$ & $0.76 \pm 0.08$ & $2.49 \pm 0.23$ & $2.82 \pm 0.33$ \\
M6 & $0.70 \pm 0.19$ & $0.81 \pm 0.15$ & $3.10 \pm 0.50$ & $3.88 \pm 1.43$ \\ \hline
M0-M2 & $0.58 \pm 0.04$ & $0.70 \pm 0.03$ & $1.59 \pm 0.09$ & $1.73 \pm 0.09$ \\
M3-M6 & $0.62 \pm 0.03$ & $0.73 \pm 0.02$ & $2.32 \pm 0.07$ & $2.60 \pm 0.09$ \\ \hline
 & \multicolumn{4}{c}{Energy} \\ \hline
M0 & \multicolumn{2}{c}{$0.49 \pm 0.07$} & \multicolumn{2}{c}{$14.77 \pm 2.26$} \\
M1 & \multicolumn{2}{c}{$0.41 \pm 0.06$} & \multicolumn{2}{c}{$12.77 \pm 1.75$} \\
M2 & \multicolumn{2}{c}{$0.41 \pm 0.04$} & \multicolumn{2}{c}{$13.07 \pm 1.12$} \\
M3 & \multicolumn{2}{c}{$0.52 \pm 0.03$} & \multicolumn{2}{c}{$16.62 \pm 0.95$} \\
M4 & \multicolumn{2}{c}{$0.48 \pm 0.03$} & \multicolumn{2}{c}{$15.34 \pm 1.00$} \\
M5 & \multicolumn{2}{c}{$0.53 \pm 0.07$} & \multicolumn{2}{c}{$16.70 \pm 2.01$} \\
M6 & \multicolumn{2}{c}{$0.41 \pm 0.10$} & \multicolumn{2}{c}{$13.49 \pm 3.06$} \\ \hline
M0-M2 & \multicolumn{2}{c}{$0.44 \pm 0.03$} & \multicolumn{2}{c}{$13.70 \pm 0.89$} \\
M3-M6 & \multicolumn{2}{c}{$0.50 \pm 0.02$} & \multicolumn{2}{c}{$15.93 \pm 0.63$} \\ \hline
\end{tabular}
\caption{Fitted parameters of the flare frequency distributions. Fitted $\alpha$ and C for the equivalent duration FFD are corrected as specified in Appendix \ref{appendix: injection tests}. We note that $\alpha$ and log C are highly correlated, and hence, so are their errors.}
\label{tab: FFD fits}
\end{table}

\subsection{Long Baseline Activity} \label{long baseline ativity}

In this section, we study the variations with respect to quiescence in our data with the goal of estimating the fraction of the total stellar energy output that can be attributed to increased activity. We do not try to identify or distinguish individual flares but rather study the entirety of the observations holistically. This approach has the advantage of not being restricted by exposure gaps, while still being of great use in examining the long term effects of NUV flares on exoplanet atmospheres and habitability. This analysis is conducted separately for each M-dwarf spectral type, wherein we aggregate data for all stars within a category.

For each individual star, we bin observed data into 300s exposure bins, which is the expected single exposure duration of \ultrasat{} \citep[see dicussion in section \ref{sec: ultrasat};]{ben-ami_scientific_2022, shvartzvald_ultrasat_2023}.
The bins are centered in each individual lightcurve, with edges being symmetrically clipped from both sides of the lightcurve to obtain an integer number of bins.
We only include sources detected in at least 3 \galex{} exposures in the subsequent analysis to ensure robust quiescent estimates. The  distribution of sources and total exposures across spectral types is shown in Figure \ref{fig: sources hist}. 
The quiescent flux for each source is obtained from the flare identification analysis detailed in section \ref{sec: flare id}.
Subsequently, each exposure bin is quiescent-normalized to enable aggregation of data from different stars. We are restricted to stars earlier than M5 in this analysis due to the paucity of data for M5 and M6 stars.

\begin{figure}
    \centering
    \includegraphics[width=\columnwidth]{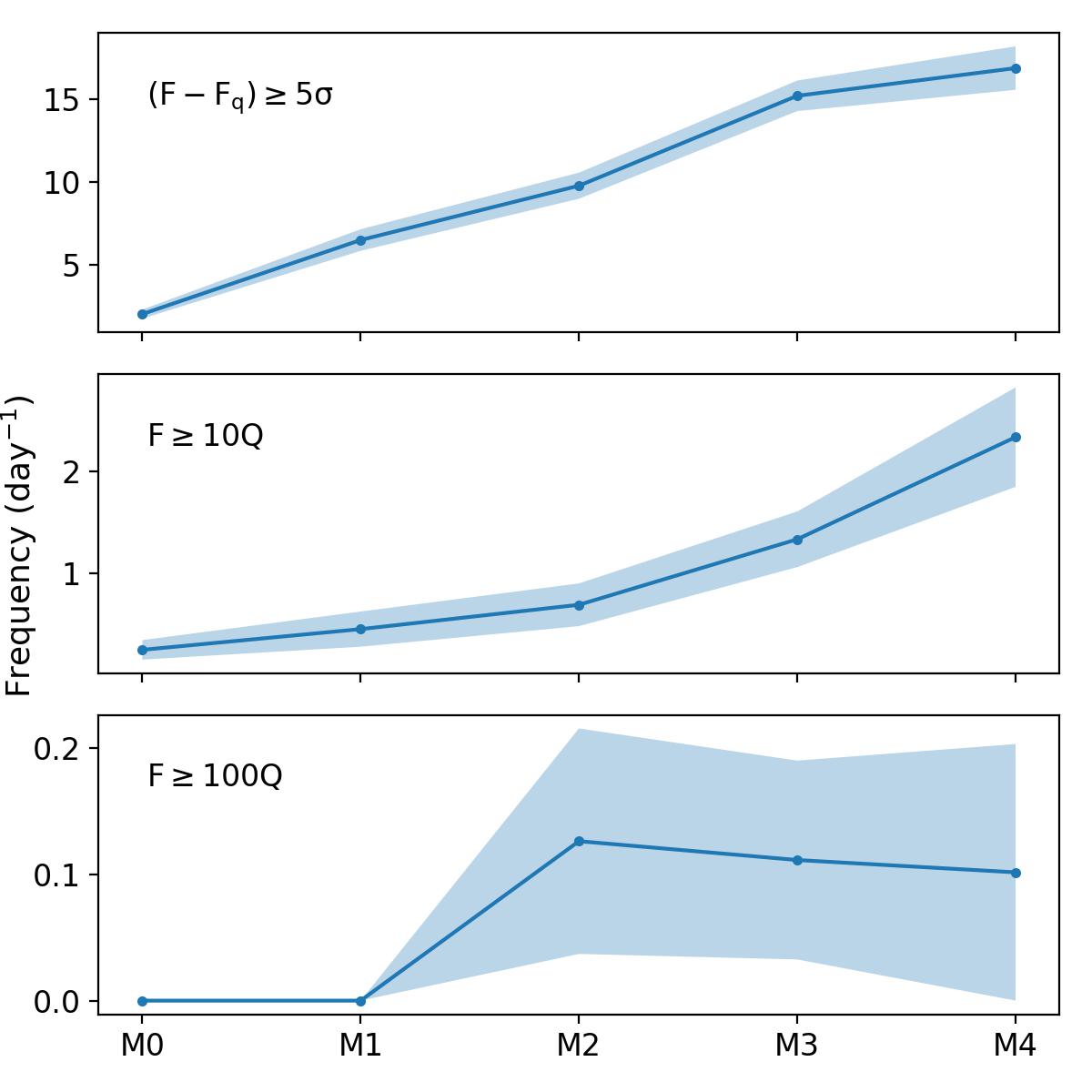}
    \caption{Occurrence rates of observed fluxes over 300s greater than $5\sigma$ above quiescent (top), 10 times quiescent (middle) and 100 times the quiescent (bottom) as a function of spectral types. We consider flux $5\sigma$ above the quiescent over a period of 300s to be a clear indication of high activity.}
    \label{fig: LTV rates}
\end{figure}

\begin{figure}
    \centering
    \includegraphics[width=\columnwidth]{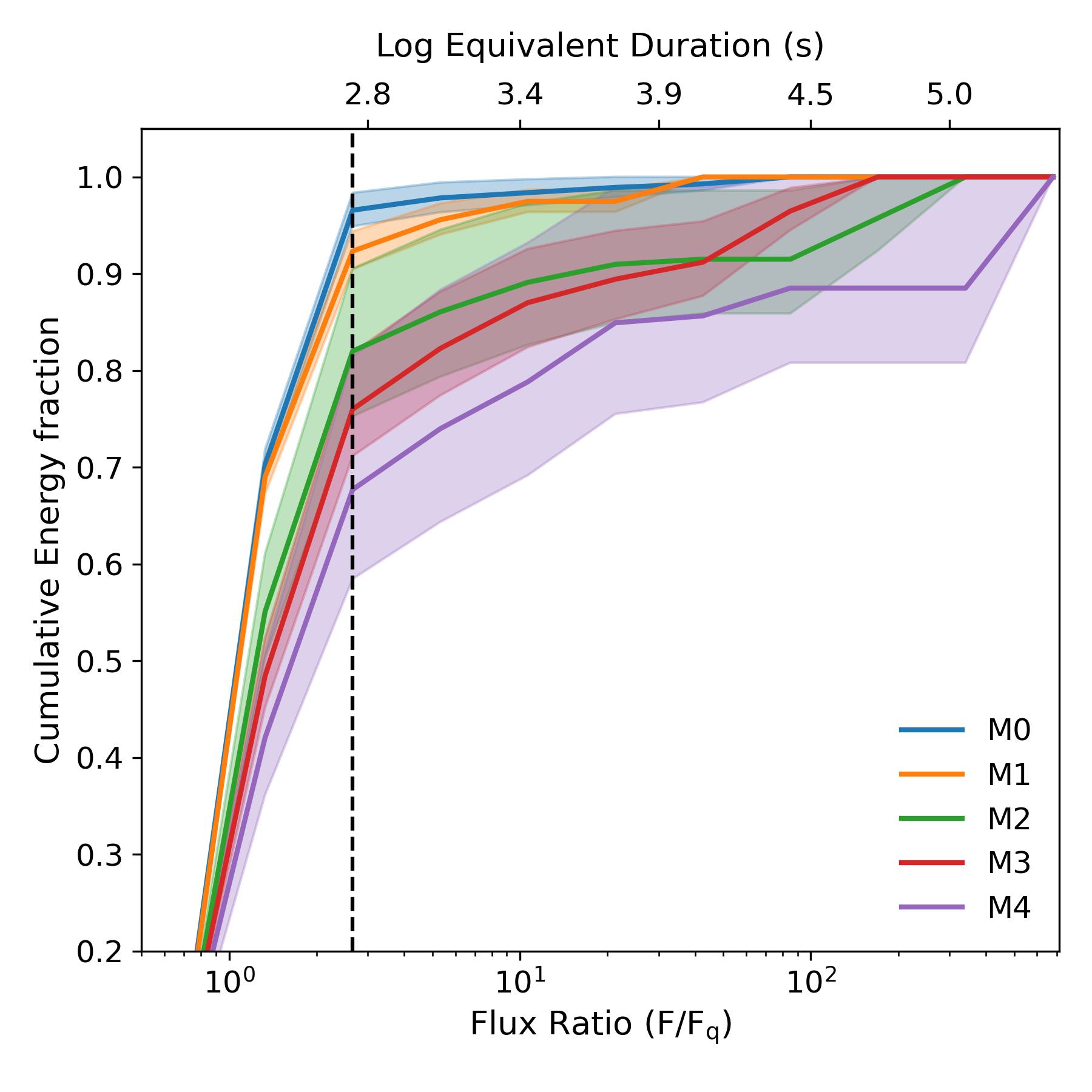}
    \caption{The cumulative fractional energy budget of M-dwarfs in the NUV as a function of local-to-quiescent flux ratios for different spectral type. The logarithm of equivalent durations corresponding to given flux ratios is shown on the top axis. The dashed line indicates flux ratio of 2.66, which is $5\sigma$ above the quiescent for a source at the SNR=3 threshold.}
    \label{fig:Cumulative LTV}
\end{figure}

The top panel in Figure \ref{fig: LTV rates} shows the occurrence rate of events $5\sigma$ above the quiescent as a function of spectral type, assuming the flux of given source to be normally distributed around the quiescent with the median error estimate as the standard deviation. The next two panels show the frequency of events more than 10 and 100 times the quiescent, which can only be caused by large flares. Occurrence rates of these events are consistent with the values predicted by the equivalent duration FFDs presented in the previous section. Figure \ref{fig:Cumulative LTV} shows the fractional cumulative energy of the stars of each category plotted as a function of the quiescent-normalized flux. With the threshold SNR being 3, flux values more than 2.66 times the quiescent would be at least $5\sigma$ above the quiescent, which we consider as increased activity. Thus we can see a clear increase in stellar activity with spectral type, in frequency as well as fractional energy output.

\section{Results and Discussion} \label{sec: Discussion}

We present the flare frequency distributions in energy and equivalent duration obtained from \galex{} data, and comparisons to previous studies in the visible/NIR and FUV bands in Figures \ref{fig: Eq dur FFD} through \ref{fig: energy ffd comparison}. Figures \ref{fig: LTV rates} and \ref{fig:Cumulative LTV} show occurrence rates of flares or enhanced activity and their contributions to M-dwarf energy budgets. We discuss the conclusions drawn from these results and their implications in the following section.

We find flaring rates to increase gradually from M0 to M6, as a function of both equivalent duration as well as energy, as seen from Figures \ref{subfig: indiv ffd delta} and \ref{subfig: indiv ffd energy}. FFDs in equivalent duration are well demarcated, with nearly 2 OOM increase from M0 to M6. The fraction of the stellar NUV energy budget attributed to flares also increases with spectral type, from $\sim 5\%$ at M0 to $\sim 33\%$ at M4. We find no evidence for an abrupt break in flaring activity at M3 or M4, where M-dwarfs become fully convective \citep{chabrier_structure_1997, baraffe_closer_2018}. 

\cite{loyd_muscles_2018} obtained FFDs for M-dwarfs in the FUV using Hubble data, albeit with a limited sample of 10 M-dwarfs. As 70\% of their sample consists of M3 and later stars, we find a general agreement between flare rates in the FUV and NUV in equivalent duration (Figure \ref{fig: compare delta FFDs}). \cite{loyd_muscles_2018} and \cite{parke_loyd_hazmat_2018} grouped their dataset into active/inactive and young/field stars respectively, but obtained no appreciable difference in flare rates of the two groups as a function of equivalent duration. Combined with our results, this may point to an exclusive link between UV flare rates (in equivalent duration) and the mechanism of the internal stellar dynamo, with other factors (e.g., rotation) playing a less significant role (see also \citealt{brasseur_short-duration_2019}).

It is significant to note that flare rates in the UV are orders of magnitude higher than those observed by TESS and Kepler in the visible/NIR \citep[white-light flares;][]{stelzer_flares_2022, althukair_main-sequence_2022}, both in equivalent duration and energy (Figures \ref{fig: compare delta FFDs} and \ref{fig: energy ffd comparison}). This implies that UV flares of M-dwarfs are either rarely accompanied by emission in the visible/NIR bands and/or the relative emission in those bands is far lesser than in the UV. 
Simultaneous observations of a flare in the Kepler and U bands by \cite{hawley_kepler_2014} found a 2 OOM suppression in relative flare amplitude in the Kepler band as compared to the U band, while a study of solar X-ray flares by \cite{watanabe_characteristics_2017} did not observe white-light counterparts for half their flare sample, supporting this hypothesis.
These results also add evidence that the canonical modelling of flare spectra solely as a $\sim 9000$K blackbody emission \citep{shibayama_superflares_2013, yang_flare_2019, gunther_stellar_2020} is inaccurate for UV flares, especially on stars post M2, as this does not produce the observed suppression in flare rates as a function of energy and equivalent duration from the UV to the visible band \citep[see also][]{kowalski_near-ultraviolet_2019, jackman_extending_2022}. We hence conclude that estimating UV flare rates and fluxes using white-light flare data is highly prone to errors.

We find 10 flares having energy greater than $10^{33}$ ergs in the NUV band, with one flare being over $10^{34}$ ergs shown in Figure \ref{fig: nonlinear flares}. These are some of the most energetic flares observed in the UV from M-dwarfs. None of the stellar hosts associated with these flares is a previously known active star (e.g., similar to AD Leonis), and we found no relevant spectroscopic data for any of them.
For reference, the canonical most energetic flare on M-dwarfs, the 1985 Great Flare on AD Leonis had an estimated bolometric energy of $10^{34}$ ergs \citep{hawley_great_1991}. The presence of such highly energetic flares can have a severe impact on planetary atmospheres and habitability, as we discuss next, and further emphasise the need of long term observations to constrain their occurrence frequency.

\subsection{Planetary Atmospheres and Habitability}

Habitable zones of M-dwarfs are closer to their host stars due to the host lower quiescent luminosity, with orbital radii ranging from $\sim0.3\,$au for M0 stars to $\sim 0.03\,$au for M6 stars \citep{segura_biosignatures_2005}. UV radiation, and consequently UV flares, are highly significant for habitability, as it is assumed UV radiation plays an essential role in the formation of life \citep[abiogenesis;][]{mulkidjanian_survival_2003, sarker_photo-alteration_2013, ranjan_influence_2016}, while on the other hand it can result in a sterilizing effect on complex life forms \citep{sagan_ultraviolet_1973}. Enhanced UV irradiance can also significantly affect concentrations of molecules associated with life on Earth (\O2, \O3, \H2O, C\O2) \citep{p_loyd_muscles_2016, meadows_exoplanet_2018, schwieterman_exoplanet_2018}, raising the possibility of biosignature false positives by future spectroscopic surveys looking for habitable planets orbiting M-dwarfs.
We focus here on two studies discussing the role of UV radiation and flares in the synthesis of molecules essential for abiogenesis: \citep{ranjan_surface_2017} and \citep{rimmer_origin_2018}, as well as work by \cite{tilley_modeling_2019} on estimating the effects of UV flares on the \O3 column density of an Earth-like planet. 

\begin{table*}[]
\centering
\begin{tabularx}{0.82\textwidth}{ccccccc}
\hline \hline
SpT & $L_{NUV}$ & Orbital Radius & NUV TOA Irradiance & $\text{F}_\text{flare}/F_q$ & $\delta$ & $\nu (\delta)$ \\
 & $\times 10^{27}$\,ergs/s & AU & ergs s$^{-1}$ cm$^{-2}$ &  & $\times 10^{3}$\,s & day$^{-1}$ \\ \hline
M0 & 12.0 & 0.277 & 55.5 & 47.3 & 250.9 & 0.002 \\
M1 & 6.7 & 0.213 & 52.4 & 50.2 & 282.9 & 0.005 \\
M2 & 3.7 & 0.179 & 40.7 & 64.8 & 478.6 & 0.015 \\
M3 & 2.8 & 0.134 & 54.6 & 48.1 & 259.6 & 0.032 \\
M4 & 2.6 & 0.090 & 113.1 & 22.7 & 55.8 & 0.166 \\
M5 & 1.4 & 0.058 & 148.6 & 17.0 & 31.0 & 0.262 \\
M6 & 0.6 & 0.034 & 178.8 & 14.0 & 20.7 & 2.554 \\
Young Sun & 7539 & 1 & 2680.7 & - & - & - \\ \hline \hline
\end{tabularx}
\caption{Top-of-atmosphere (TOA) irradiances for habitable zone planets and required relative flare amplitudes and corresponding equivalent durations and occurrence rates to reach young Sun and Earth TOA irradiance. Luminosities are weighted averages, with weights equal to the inverse of each star’s accessible volume. The young Sun NUV luminosity was determined from the spectrum given in \cite{ranjan_surface_2017}. Orbital radii are obtained from the formula given by \cite{ranjan_surface_2017} and the bolometric luminosities given by \cite{pecaut_intrinsic_2013}. Equivalent durations corresponding to relative flare amplitudes are obtained using the equivalent duration - flare length relation determined in Eq. \ref{eqn: rel amp - eqd} and Fig. \ref{fig:D vs EqD}.}
\label{tab: ranjan_table}
\end{table*}

Using archival M-dwarf UV spectral and photometric observations, \cite{ranjan_surface_2017} investigated the beneficial as well as detrimental effects of NUV radiation on fundamental photochemical reactions on a young Earth-like planet orbiting M-dwarfs in their habitable zone. They found reaction rate ratios of beneficial to detrimental prebiotic processes (eustressors and stressors) to be similar to those on young Earth, but reaction rates having a suppression of 2-4 OOM due to a similar suppression in the planetary NUV irradiance of M-dwarfs as compared to a young Earth and young Sun. \cite{ranjan_surface_2017} suggested that an ersatz day-night cycle caused by regular occurrences of large flares may be sufficient for enhanced rates of abiogenesis. Similarly, \cite{rimmer_origin_2018} estimated surface NUV irradiance levels required in the band 200-280 nm to obtain sufficient production of RNA pyrimidine nucleotides, an important building block of life.


Table \ref{tab: ranjan_table} shows the average NUV top-of-atmosphere (TOA) irradiances for planets in the habitable zones of M0 to M6 dwarfs in our \galex{} sample, as well as threshold flare amplitudes and corresponding frequencies to reach the TOA NUV irradiance of the young Sun. To correct for Malmquist bias, we compute the luminosities for each spectral type as weighted averages, where each measurement is weighted by the inverse of the maximum volume over which it could be detected.
We find NUV TOA irradiances to be around 20-50 times (1.5 OOM) lower than the Young Sun, with flares reaching the young Earth irradiance level occurring around once every 5 days for stars later than M3 (except for M6 stars which however are reliant on very small sample statistics).
On the other hand, with the NUV luminosities determined from \galex{} data, and assuming minimal atmospheric attenuation in the 200-280 nm band \citep[see ][]{}\cite{ranjan_surface_2017}, we find surface NUV irradiances to be within the range for RNA production for all M-dwarfs as per the work of \citet{rimmer_origin_2018}\footnote{Following correspondence with P. Rimmer.}. These results imply that TOA NUV fluxes seem to be at the threshold for the emergence of life for M-Dwarfs later than M3, and point to the need of robust laboratory experiments needed to pin down prebiotic reaction rates dependent on flaring activity.

\begin{figure}
    \centering
    \includegraphics[width=0.9\columnwidth]{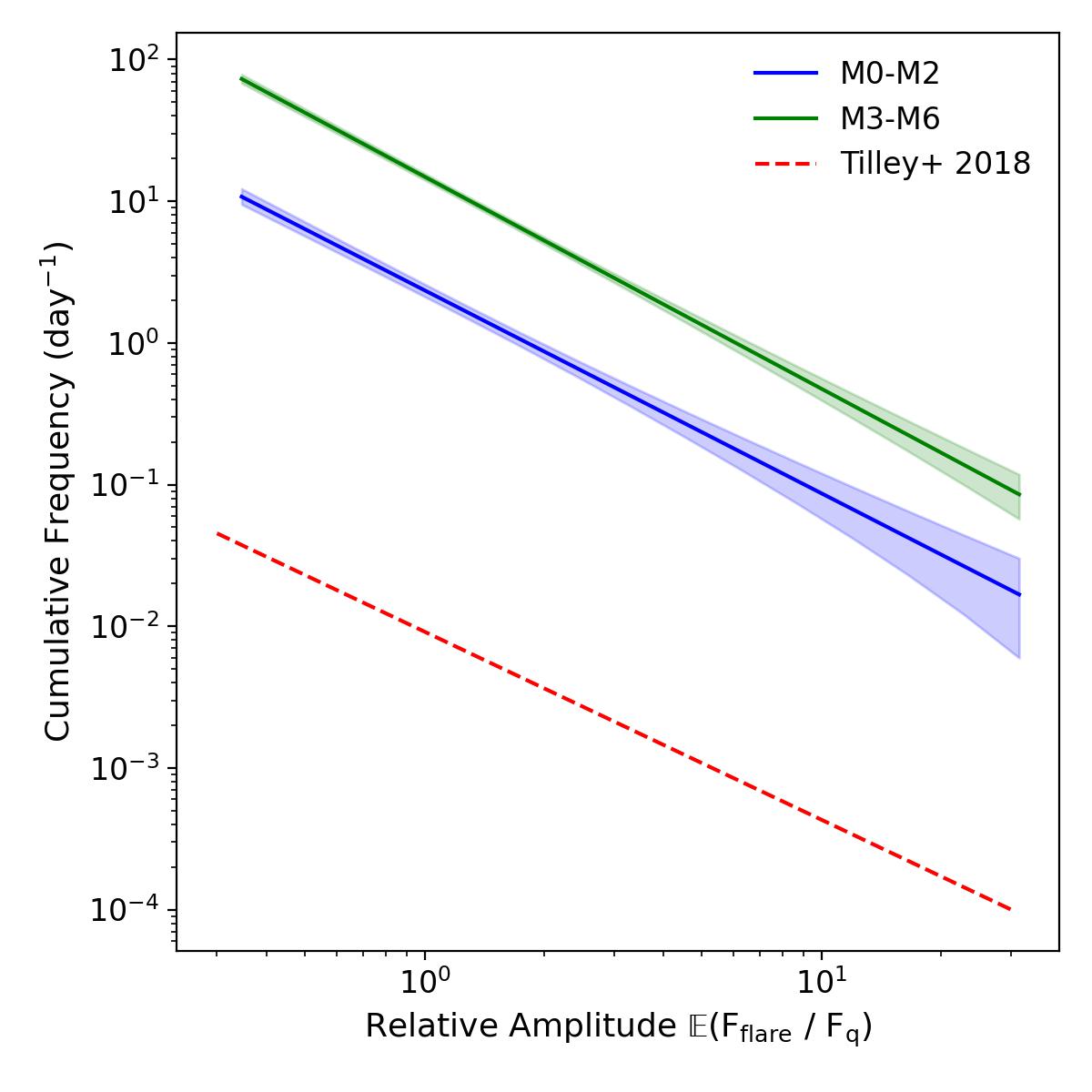}
    \caption{Comparison of flare relative amplitude frequency distributions used by \cite{tilley_modeling_2019} to those estimated from \galex{} data. \galex{} relative amplitude frequency distributions are obtained using the equivalent duration - flare length relation determined in Eq. \ref{eqn: rel amp - eqd} and Fig. \ref{fig:D vs EqD}. }
    \label{fig: tilley relamp}
\end{figure}

The atmospheric model used by \cite{ranjan_surface_2017} does not have ozone, as the ozone layer on Earth is understood to have been formed due to biotic activities roughly 800 million years ago \citep{ligrone_great_2019, cooke_revised_2022}. While NUV radiation may be essential for abiogenesis, it is harmful to complex organisms, which are shielded by the ozone layer on Earth. Simulations of planetary atmospheres of unmagnetized Earth-like exoplanets by \cite{tilley_modeling_2019} have raised the possibility of ozone columns being partially or completely depleted due to coronal mass ejections (CMEs) accompanying flares. \cite{tilley_modeling_2019} utilized a U-band FFD derived from Kepler observations of an M4V star along with Kepler to U-band scaling \citep{hawley_kepler_2014}, correlating U-band relative flare amplitudes to observations of our Sun to derive CME rates and proton fluxes. 
The FFD in energy used by \cite{tilley_modeling_2019}, shown in Figure \ref{fig: energy ffd comparison}, is lower than our \galex{} FFDs by 1-3 OOM from early to late spectral M-dwarf spectral types at $10^{34}$ ergs.
Furthermore, we find that the flare relative amplitudes used by \cite{tilley_modeling_2019} are more than 3 OOM lower than ones determined from \galex{} observations as shown in Figure \ref{fig: tilley relamp}. As CME proton fluxes are proportional to flare amplitudes \citep[Eq. 5,][]{tilley_modeling_2019}, this indicates 3 OOM higher proton fluxes impinging upon planetary atmospheres than those used by \cite{tilley_modeling_2019} in their simulations.
The combined effect of these observations suggest that habitable zone planets would not be able to build up substantial ozone layers, allowing sterilizing levels of NUV radiation to reach the planetary surface. 

Our results thus paint the following picture: flares might allow exoplanets orbiting M4 and later dwarfs to have sufficient surface NUV flux for abiogenesis. Nevertheless, those same flares may preclude the formation of an ozone layer, exposing subsequent complex lifeforms to sterilizing levels of UV radiation. For instance, the threshold NUV dosage for 10\% survival of the common bacterium \textit{E. coli} is $2.2\times10^4$ ergs/cm$^2$ \citep{estrela_superflare_2018, estrela_chapter_2021}, which is reached within minutes for habitable-zone exoplanets orbiting any M-dwarf without attenuation from the ozone layer. 
However, life could still survive in locations shielded from UV radiation, such as underwater or underground. Continuing the previous example, \textit{E. coli} would survive at a depth 10-30\,m underwater even during the strongest flares we observe, which is well within the photic zone (depth < 200\,m) .

\begin{figure}
    \centering
    \includegraphics[width=0.85\columnwidth]{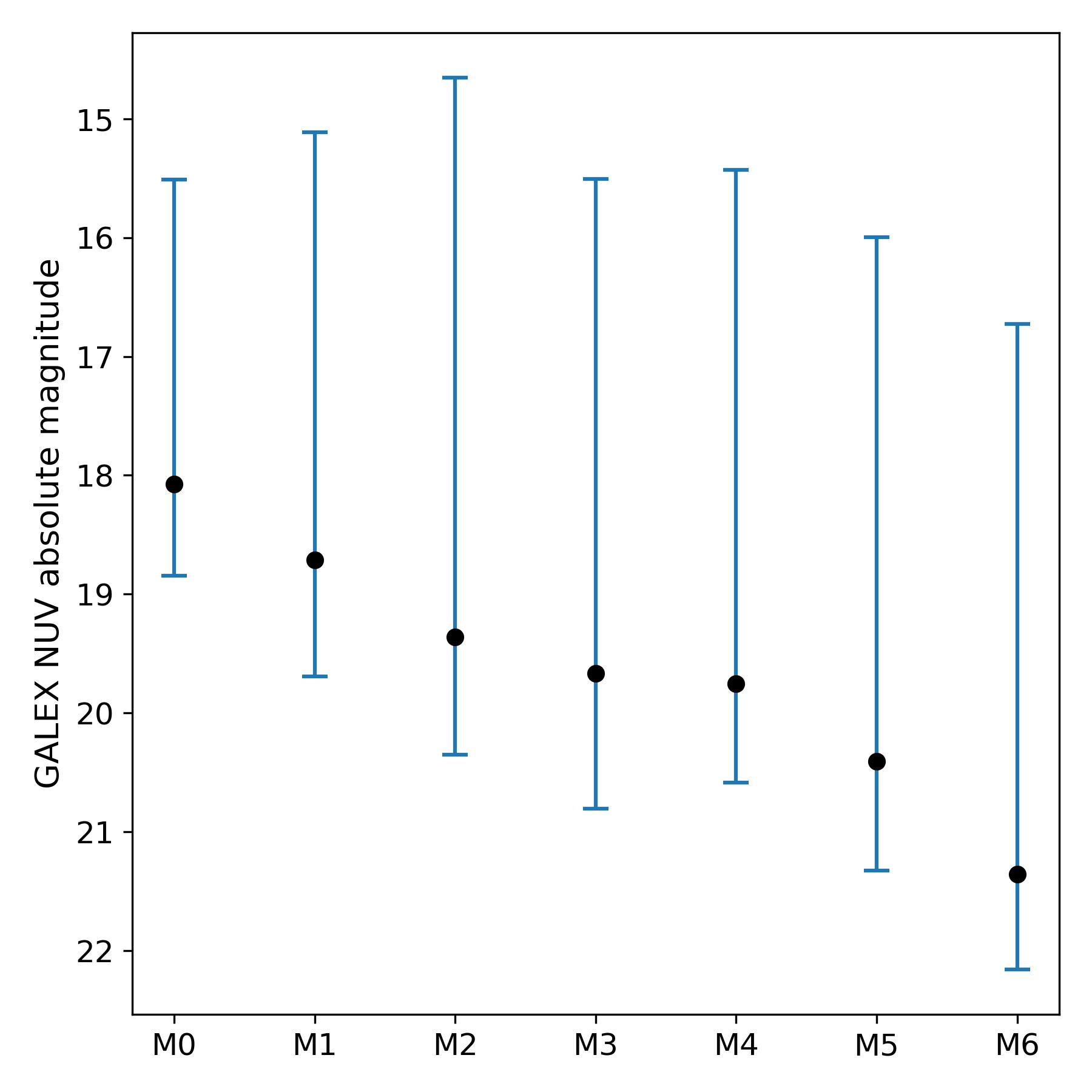}
    \caption{Volume-corrected mean NUV M-dwarf quiescent absolute magnitudes (black) along with unweighted central 95\% ranges (blue).}
    \label{fig: galex nuv mags}
\end{figure}
 
Even as the large size of the \galex{} dataset enables predictions about UV flares and their effect on exoplanetary habitability to far greater accuracy than previously possible, it pays to keep in mind that that these results are based on averaged NUV luminosities and flaring activity levels.
Quiescent NUV luminosities of M-dwarfs have large variability, even within the same spectral sub-type (Figure \ref{fig: galex nuv mags}; see also \citealt{miles_hazmat_2017, cifuentes_carmenes_2020}), and flare rates and activity levels may vary among stars as well. This is especially true for young M-dwarfs, which have higher quiescent NUV luminosities and activity levels. Furthermore, flare rates at at high equivalent durations and energies ($>10^5$\,s and $>10^{34}$\, ergs) are supported by a sample size of 4 and 1 respectively, due to the rarity of these events and the short durations of \galex{} exposures. 
We thus emphasise the need for extensive and long-term observations of M-dwarfs in the UV, which would enable the characterization of UV levels and flare rates for M-dwarfs belonging to different populations (as a function of age, metallicity, rotational velocity etc.) and create a larger sample of highly energetic flares. The \ultrasat{} mission would enable the creation of such a flare catalog, as discussed in the next section. Furthermore, focused photometric and spectroscopic observations of M-dwarfs which are of high interest as exoplanetary hosts (e.g. Proxima Centauri and TRAPPIST-1) may be particularly helpful. Such observations can be provided by upcoming missions such as SPARCS and UV-SCOPE.

\section{Future observations with ULTRASAT} \label{sec: ultrasat}
ULTRASAT is a scientific satellite equipped an NUV telescope ($230-290\,$nm) with a FoV of $\sim$ 170 deg$^2$. It is led by the Weizmann Institute of Science and the Israel space agency in collaboration with DESY (Helmholtz association, Germany) and NASA (USA). ULTRASAT is expected to be launched to geostationary transfer orbit (GTO) in 2027, and will operate at geostationary orbit (GEO). The mission targets time-domain astrophysics phenomena, and is the first wide-FoV telescope operating in the NUV band. For sources with T$\sim20,000\,$K, ULTRASAT is comparable in grasp - the volume of space probed per unit time \citep{ofek_seeing-limited_2020} - to that of the Vera C. Rubin observatory \citep{ivezic_lsst_2019}. ULTRASAT exposures are set to $300\,$s. For a single exposure, ULTRASAT is expected to have a limiting magnitude of $22.0$ for a blackbody spectrum of $>10,000$K, while for cool sources such as M-dwarfs, the estimated limiting magnitude is $20.8$. The co-add limiting magnitude for cool dwarfs is estimated to be $23.5$. ULTRASAT will conduct several survey modes, with the majority of observing time dedicated to high cadence continuous observations of one pre-selected field at a high galactic latitude\footnote{This is preliminary, and observing strategy might change by the time the mission is operational.}.

One of the science cases \ultrasat{} will address is the frequency of high-energy flares / episodes of activity for M-dwarfs. We further note that accurate classification of the large numbers of M-dwarf flares as such is crucial to avoid false positive alerts for events such as supernovae shock-breakouts. A thorough mapping of the \ultrasat{} high cadence fields in visible/NIR bands, as well as observing strategies with visit lengths longer than the expected impulse phase of a flare will guarantee the number of false positives being kept at a minimum. We turn now to estimate the number of M-dwarfs sources detected in the \ultrasat{} FoV due to flares, which would otherwise be too faint for \ultrasat{} to detect in quiescence given these sources' low UV brightness. We assume UV flares would have a bluer (\textit{i.e., }hotter) spectrum than during quiescence, approximating it to a blackbody spectrum of $\geq$10,000\,K.   

We run a simulation to estimate the numbers of detectable M-dwarf sources within a single $300\,$s exposure (located at galactic latitude 47.6; see Appendix \ref{appendix: G-M estimation}), divided into two categories: early M-dwarfs (M0-M2) and mid-M-dwarfs (M3-M6). Each iteration of the simulation generates a sample of M-dwarfs of both categories along with their distances and corresponding apparent magnitudes. A challenge faced in this task is the incompleteness of the known M-dwarf sample in large visible/NIR band surveys beyond $\sim200\,$pc, depending upon spectral type. This is due to M-dwarf low brightness, as well as the breakdown of standard color-temperature relations below 3,000\,K. While M-dwarfs located further than these distances would be undetectable by \ultrasat{} in quiescence, it is imperative to estimate their population up to farther distances to estimate the occurrence rate of M-dwarfs detectable through flares. We achieve this by tying the number density of M-dwarfs to that of G-dwarfs to a distance of $800\,$pc; see further details in Appendix \ref{appendix: G-M estimation}. 

\begin{figure}
    \centering
    \includegraphics[width=0.9\columnwidth]{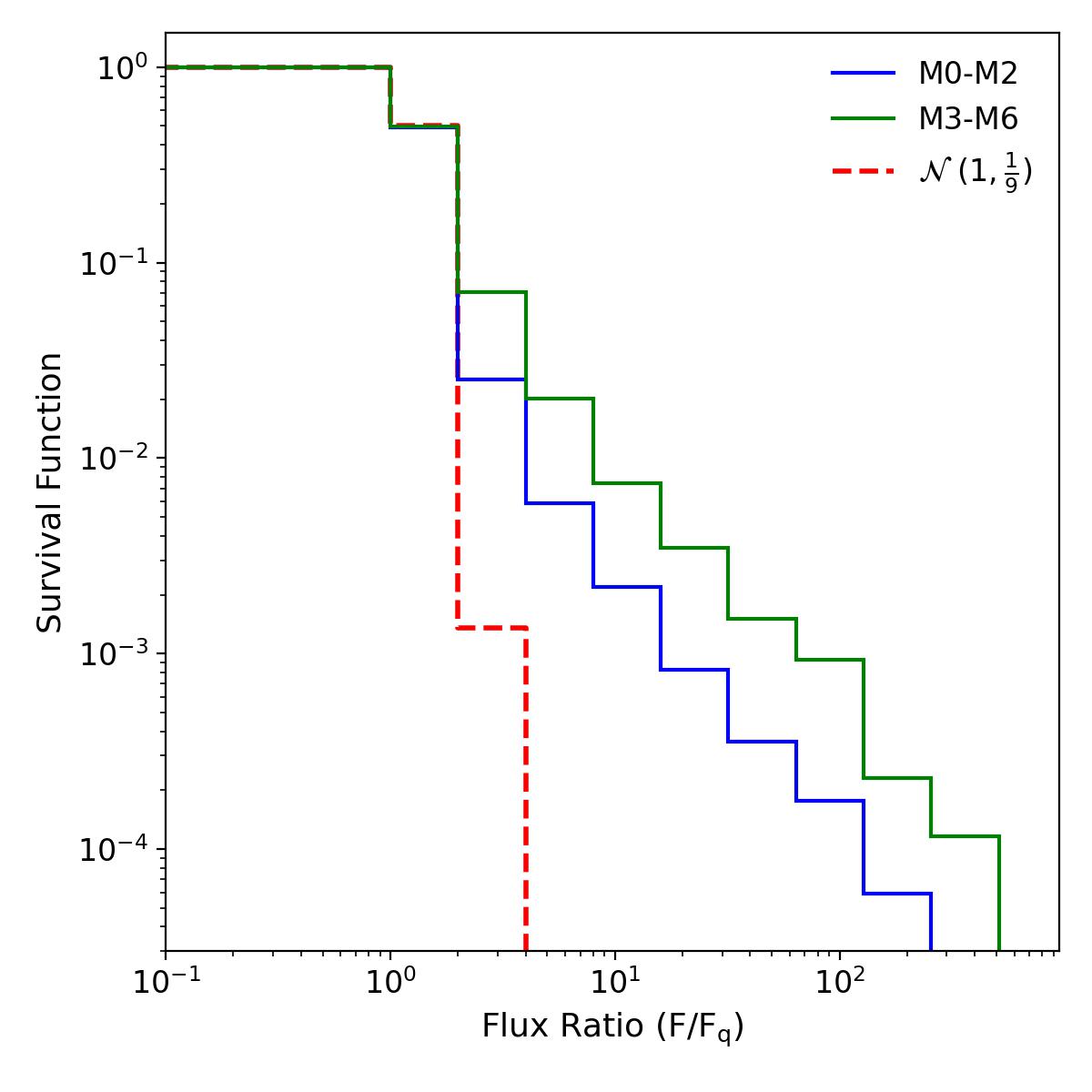}
    \caption{Cumulative probability distributions of local-to-quiescent flux ratios for M0-M2 dwarfs (blue) and M3-M6 dwarfs (green). These are used to estimate the potential number of flares detected by ULTRASAT in both categories. The normal distribution for an SNR=3 source (set as the threshold of the \galex{} sample) is shown in red.}
    \label{fig:survival_fn}
\end{figure}

Stars beyond the cool dwarf limiting magnitude of 20.8 need to be at commensurately high activity levels to be detectable, the probabilities of which are obtained from the long baseline activity study described in section \ref{long baseline ativity}. Specifically, we obtain survival functions of local-to-quiescent flux ratios for M-dwarfs in both categories, as shown in Figure \ref{fig:survival_fn}. The red dashed line in Figure \ref{fig:survival_fn} corresponds to a normal distribution for an SNR=3 source, which is the limiting threshold of our \galex{} sample. The survival function significantly deviates from the normal distribution post $3\sigma$, or $F/F_q = 2$. We consider the regime beyond this limit to be caused exclusively due to high activity or flares; hence the smallest detectable `flare' has a flux ratio of 2.

We conduct 10000 iterations and find each 300s exposure has 152 flaring M-dwarf stars detected on average, 53 being M0-M2 and 99 being M3-M6 stars. Of these, 1-2 flares occur on stars visible in quiescence, while the other $\sim 150$ flares occur on stars fainter than the limiting magnitude, thereby brightening them enough to be detected. These would add up to $\mathcal{O}(10^6)$ M-dwarf flares detected by \ultrasat{} over 6 months of continuous exposure. We find an average field at this galactic latitude would contain 36 M0-M2 stars and 9 M3-M6 stars visible in quiescence. Given the confusion limit of 24 magnitude, we further expect to determine the quiescence level of $\sim 2800$ M0-M2 stars and $\sim 1000$ M3-M6 stars in the FoV following repeated observations.

\section{Summary} \label{sec: summary}

In this work, we have aimed to constrain the flaring activity of M-dwarfs in the NUV using archival data from \galex{} and \xmm{}. The \galex{} observations form the most extensive dataset of M-dwarfs in the NUV to date, with exploitation of this data possible due to the new \gphoton2 pipeline. We identify and characterise flares and determine flare frequencies as a function of equivalent duration and energy for stars from M0 to M6 in the \galex{} dataset. We also study long baseline activity with the \galex{} dataset, determining rates of enhanced activity and flares and their contributions to stellar energy budgets for stars as late as M4.

The presented dataset allows us to constrain flare frequency distributions in the NUV up to $10^5\,$s in equivalent duration and $10^{34}$ ergs in energy, orders of magnitude above any previous study in the UV. We find flare rates increase monotonically with spectral type, with a 2 OOM increase from M0 to M6. This also increases the contribution of flares to total stellar energy budgets in the NUV, from $\sim 5\%$ at M0 to $\sim 33\%$ at M4. We find 10 flares with NUV energies above $10^{33}$ ergs, including one of the most energetic flares observed on M-dwarfs having NUV energy exceeding $4\times10^{34}$ ergs. Comparing our results to \cite{loyd_muscles_2018}, we find a general agreement between flare rates in the NUV and the FUV. We find flare rates in the UV to be several OOM above previously determined rates in the visible/NIR bands, further concluding current methods of extrapolation of flare rates/strengths in the UV from visible/NIR observations to be highly prone to error.

We obtain median quiescent NUV luminosities for M0-M6 stars and hence estimate NUV irradiance for planets in habitable zones.
Combined with estimates of flare rates, we speculate that the NUV irradiance for planets orbiting stars M4 and later stars may be on the threshold for enabling abiogeneis.
On the other hand, we find the high frequencies of energetic UV flares and associated CMEs likely inhibit the formation of an ozone layer, exposing planetary surfaces to sterilizing levels of UV radiation, thereby potentially preventing genesis of complex Earth-like lifeforms.

\galex{} data indicates large variability in NUV luminosities even within individual spectral types (Appendix \ref{appendix: galex Q}), which would have significant impact on flare energies and habitability. This is further complicated by the higher luminosities and activity levels of young M-dwarfs, with habitable zones shifting inward as the stars age. We thus require an extensive campaign of long term UV observations of M-dwarfs. These would be provided by upcoming missions like ULTRASAT. We estimate ULTRASAT to detect $\mathcal{O}(10^6)$ M-dwarf flares over 6 months of continuous exposure. When coupled with spectroscopic observations, this will enable the scientific community to form a better understanding of NUV flare activity on M-dwarfs, and its impact on atmospheric evolution and habitability.

\section*{Acknowledgements}
We thank Chase Million, Michael St. Clair, and Scott W. Fleming and the rest of the \gphoton{} team for their kind help and prompt responses to our queries. We thank the ESA XMM-Newton helpdesk for their assistance. We thank Paul Rimmer for his input on atmospheric attenuation and prebiotic chemistry. We thank Sukrit Ranjan, Kevin France, and Pat Behr for pointing out observational biases in our analysis, with special thanks to Sukrit Ranjan for his valuable insights.

S.B.A is grateful for support from the Willner family foundation, Israel Science Foundation, Israel Ministry of Science, Minerva and the Azrieli Foundation.

\facilities{GALEX, XMM, MAST}

\software{gPhoton2 v3.0.0a0 \citep{million_gphoton_2016,st_clair_gphoton2_2022},
Photutils v1.6.0 \citep{bradley_astropyphotutils_2022},
Astropy v5.2 \citep{the_astropy_collaboration_astropy_2013, the_astropy_collaboration_astropy_2018, the_astropy_collaboration_astropy_2022},
emcee v3.1.3 \citep{foreman-mackey_emcee_2013},
celerite v0.4.2 \citep{foreman-mackey_fast_2017},
brokenaxes v0.5.0
}

\appendix
\counterwithin{figure}{section}
\counterwithin{table}{section}

\section{Injection Tests} \label{appendix: injection tests}

\begin{figure}
    \centering
    \includegraphics[width=\columnwidth]{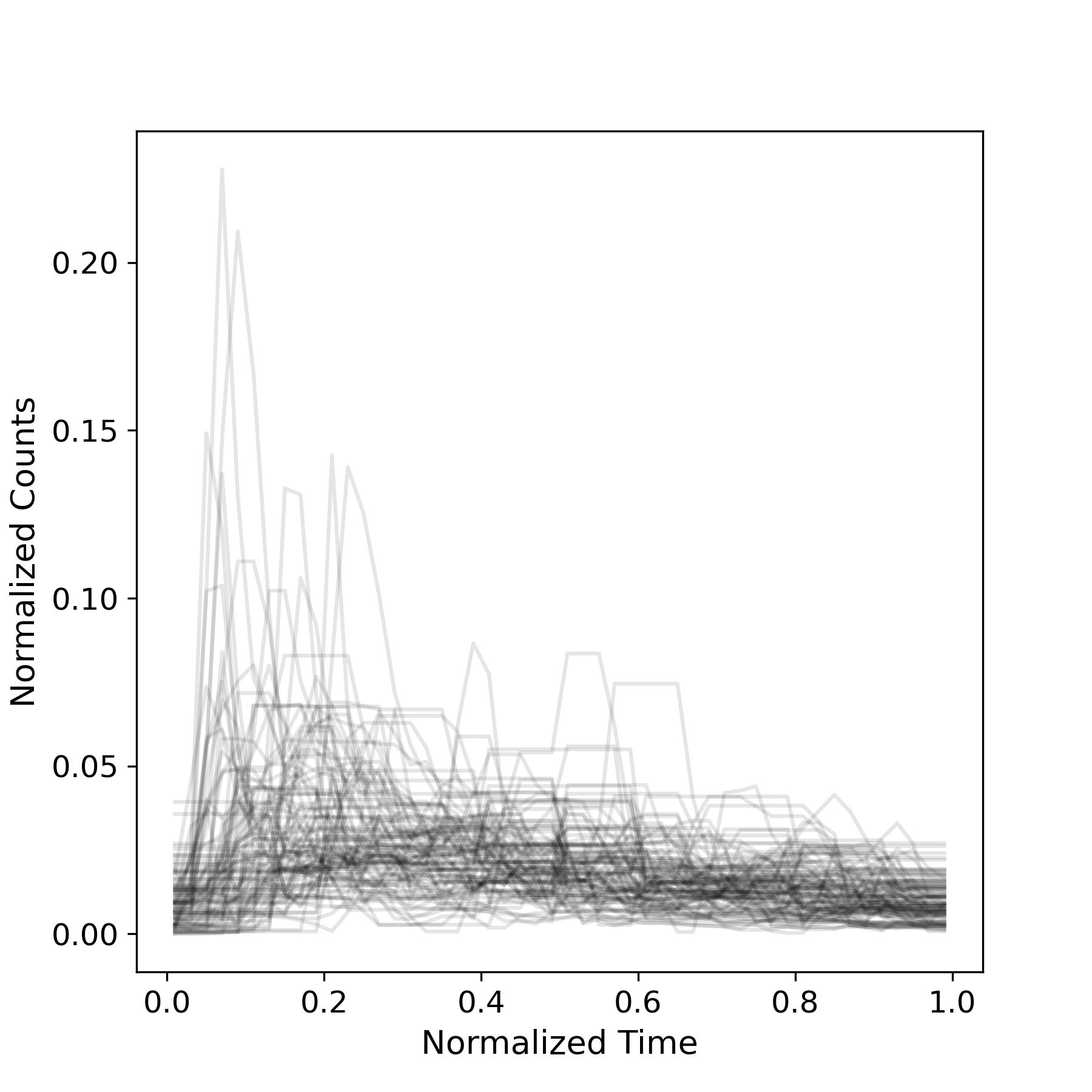}
    \caption{Superposition of all flare profiles in the profile bank.}
    \label{fig: flare profiles}
\end{figure}

\begin{figure}
    \centering
    \subfloat[]{
        \includegraphics[width=0.9\columnwidth]{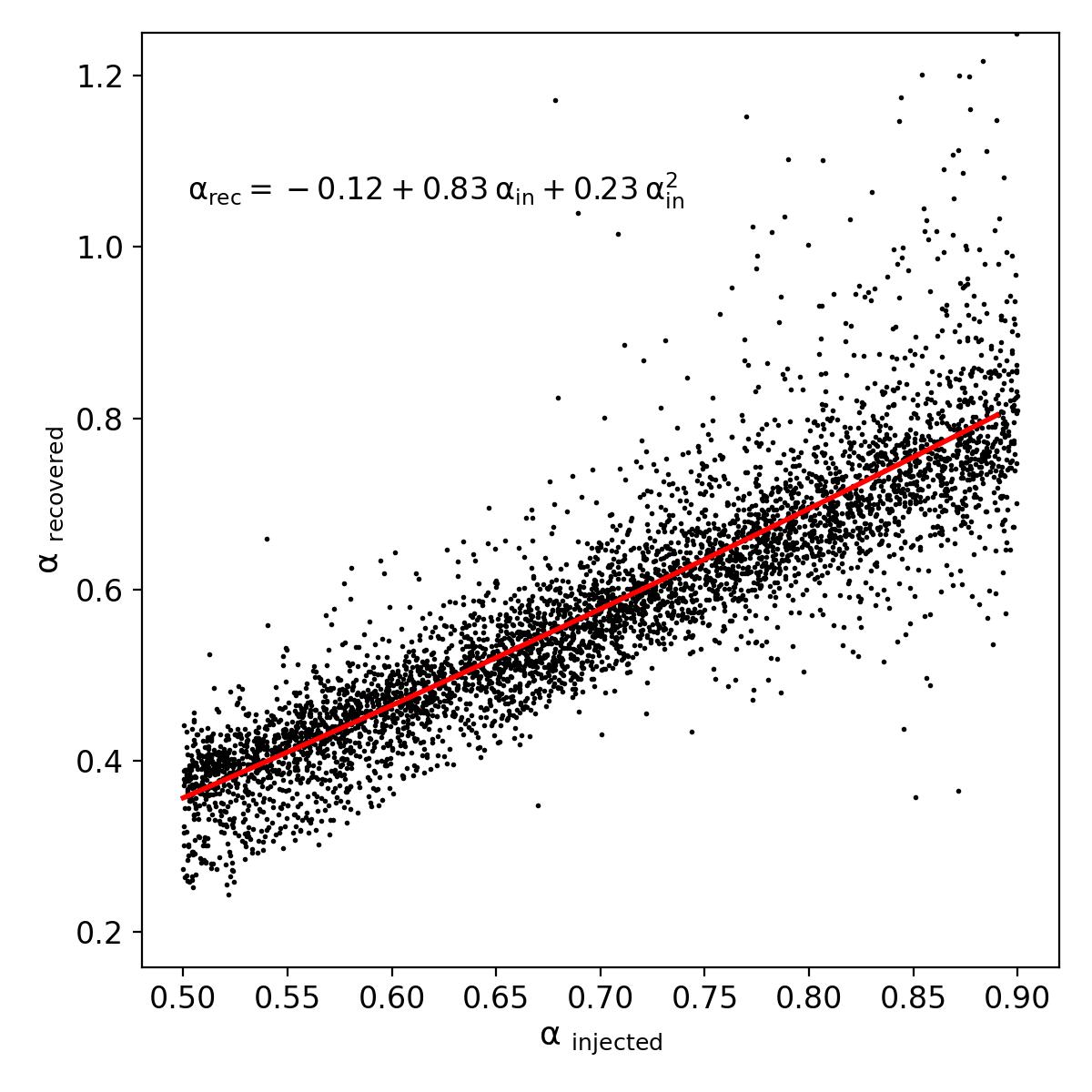}
    }\\
    \subfloat[]{
        \includegraphics[width=0.9\columnwidth]{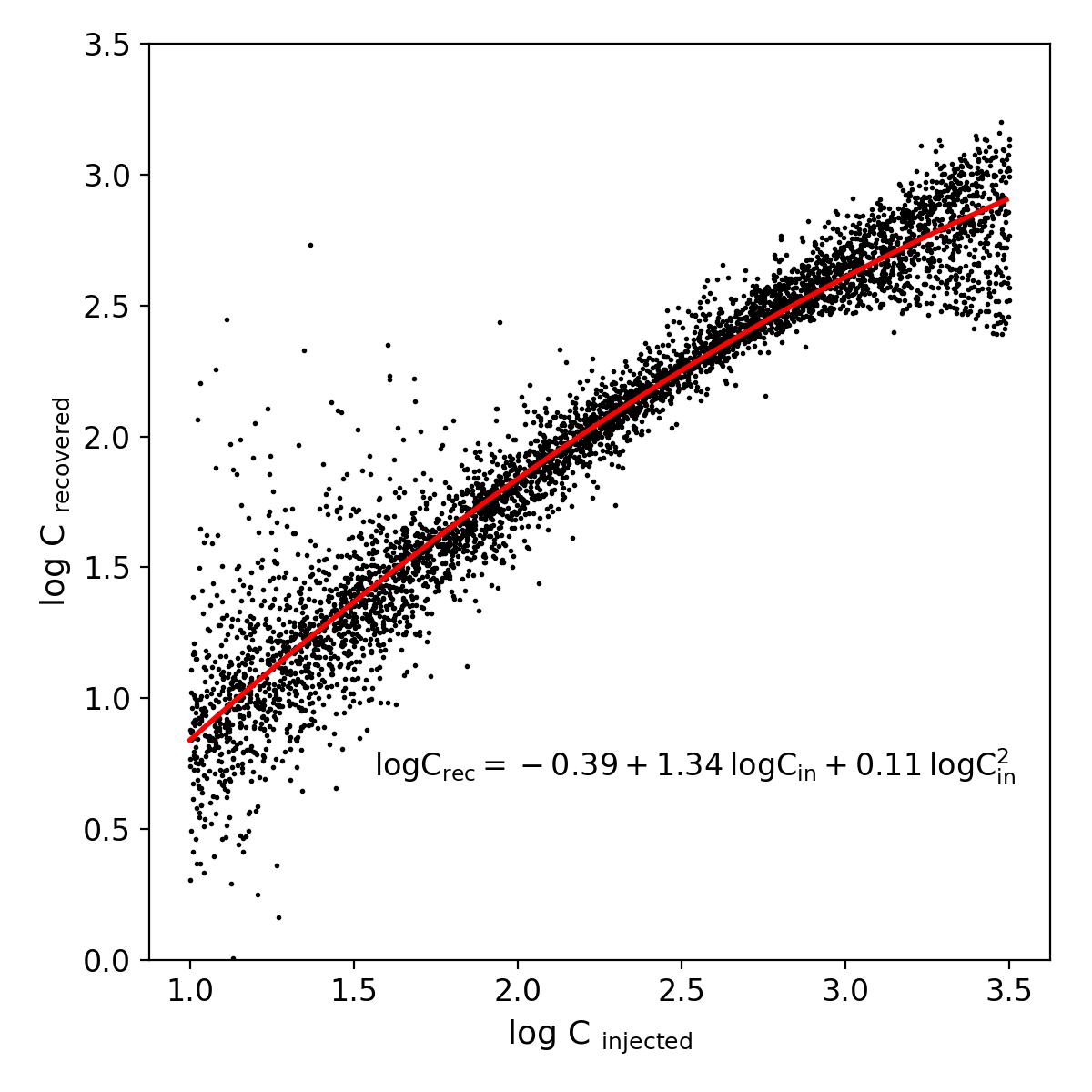}
    }
    \caption{Flare Injection Recovery simulation results for the power-law fit parameters $\alpha$ and log\,C. The red curve shows the polynomial fit for the relation between injected and recovered fit parameters. Recovered parameters are shifted from injected values due to exposure gaps inhibiting complete detection of some flares.}
    \label{fig: injection correction}
\end{figure}

We perform flare injection/recovery tests to test the accuracy of our equivalent duration FFD fits and verify the theoretical lower detection limits given in section \ref{sec: flare id}. As a first step, we remove all suspect and anomalous regions identified by the flare detection pipeline from the concatenated lightcurves. The resultant gaps are filled in with random normal data having the mean and variance of the clipped lightcurve. 
Flares are injected at random locations in each lightcurve with the flare shape given by a randomly selected profile from the flare profile bank described next.

As we observe a wide variety of flare profiles across the entire dataset, we create a bank of flare profiles using completely detected flares (as per the criteria given in section \ref{subsec: partial flares}). To reduce noise contributions in the flare profiles, and to maintain reasonable time resolution, we restrict the bank to flares with equivalent duration above 100s and covering at least 5 points of the lightcurve, detected in sources binned at a cadence of 30\,s or better. We also exclude flares having any portion of the flagged region below the quiescent. The bank consists of 111 flares, shown in Figure \ref{fig: flare profiles}. The flare profiles are then normalized to have unit fluence, and stored as linearly interpolated cumulative time profiles so as to have infinite time resolution.

In each simulation iteration, we randomly select 20 sources from the \galex{} sample, with selection probability of a source being weighted by the number of exposures in which the source is detected. Each source lightcurve is sub-iterated n times such that the total observable exposure time within an iteration is approximately 125 days.
To approximate real conditions, gaps between (as well at the start of) continuous segments in a concatenated lightcurve are uniformly randomized between 4000\,s and 12000\,s. 
The number of injected flares in each lightcurve are randomized from a Poisson distribution with $N$ given by Eq. \ref{eqn: expected flare numbers} and $\Delta T$ being the total exposure duration of a lightcurve including gaps. Their equivalent durations are randomized from the power-law distribution given by Eq. \ref{eqn: flare delta distribution}.
Injected flares have equivalent durations between $\delta_\text{low}=20\,$s and $\delta_\text{high}=10^7\,$s. $\alpha$ and log\,C are uniformly randomized in the ranges (0.5, 0.9) and (1, 3.5) respectively, which well cover the observed parameter space.
Flares are scaled in time as per the equivalent duration - flare length relation given in  Eq. \ref{eqn: eqd vs D} (Figure \ref{fig:D vs EqD}), with the power-law constant and index randomized from a normal distribution with the given mean and standard deviation.
\begin{equation}
    \text{Flare Length} = 10^{0.96\pm 0.07} \delta^{0.51\pm 0.03.}
    \label{eqn: eqd vs D}
\end{equation} 
We then run the flare detection and FFD fitting algorithms (sections \ref{sec: flare id} and \ref{sec: FFD fits}) on the simulated lightcurves to obtain fits for $\alpha$ and log\,C.

Figure \ref{fig: injection correction} plots the input and output power-law parameters for 5000 iterations. We find a polynomial relation between the injected and recovered parameters as shown in the figures. We use the inverse of these relations to correct the equivalent duration power-law fits in section \ref{sec: FFD fits}.

\begin{figure}
    \centering
    \includegraphics[width=\columnwidth]{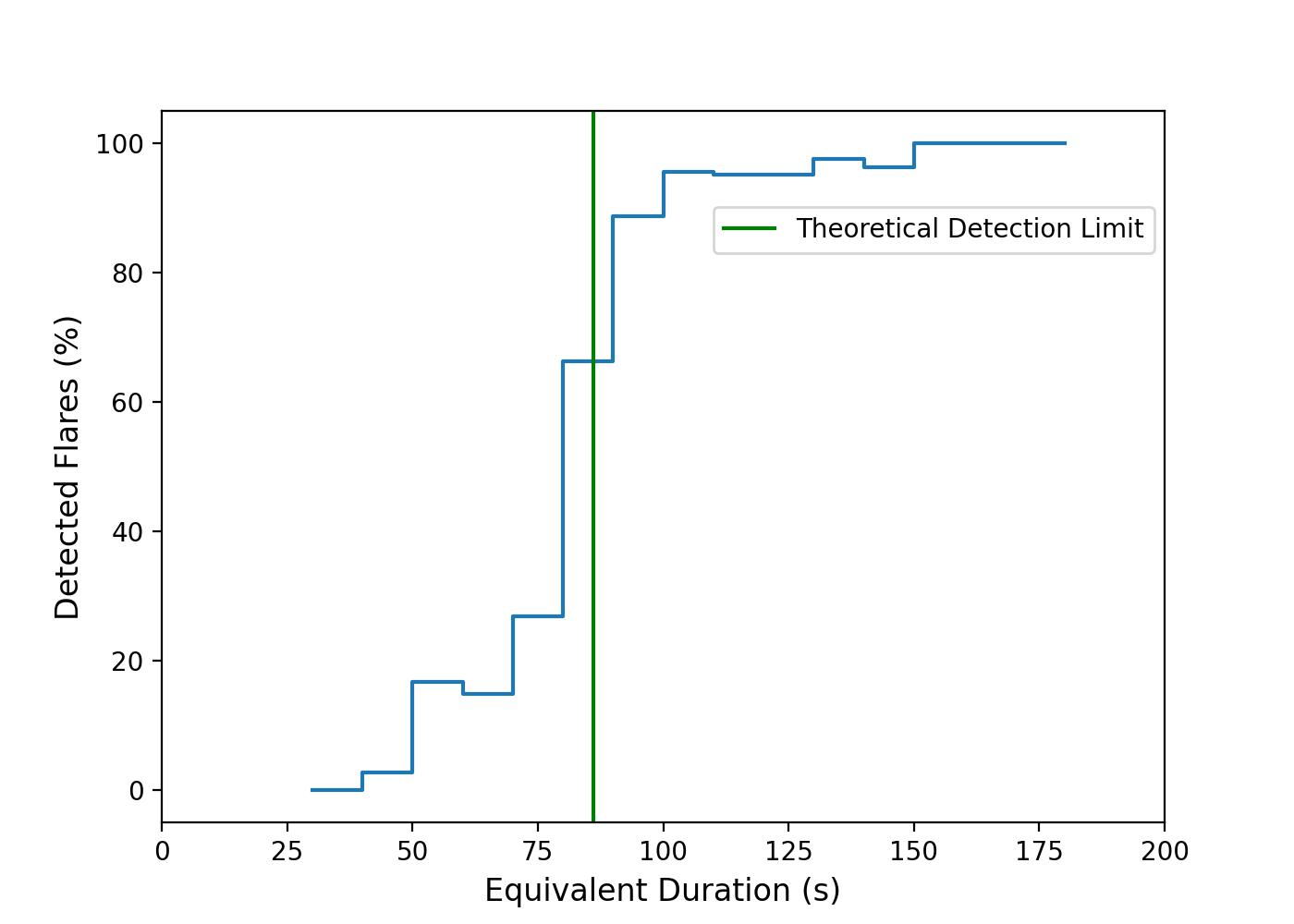}
    \caption{Example of injection/recovery simulations for a random \galex{} source. The green line shows the theoretical detection limit, which matches with the start of the recovered detection plateau.}
    \label{fig:low lim inj test}
\end{figure}

For verifying the lower limit, we inject flares with equivalent durations uniformly distributed between zero and twice the theoretical lower detection limit. Figure \ref{fig:low lim inj test} shows the result of an injection/recovery test with 5000 iterations for a randomly chosen lightcurve. The theoretical detection limit is seen to be at the beginning of the observed detection plateau, which ties with our expectations. We also used the injection/recovery trials to test for false positives, but found that those become negligible well below the detection limit. 

\section{\galex{} Quiescent Flux and Luminosities} \label{appendix: galex Q}

\begin{figure}
    \centering
    \includegraphics[width=\columnwidth]{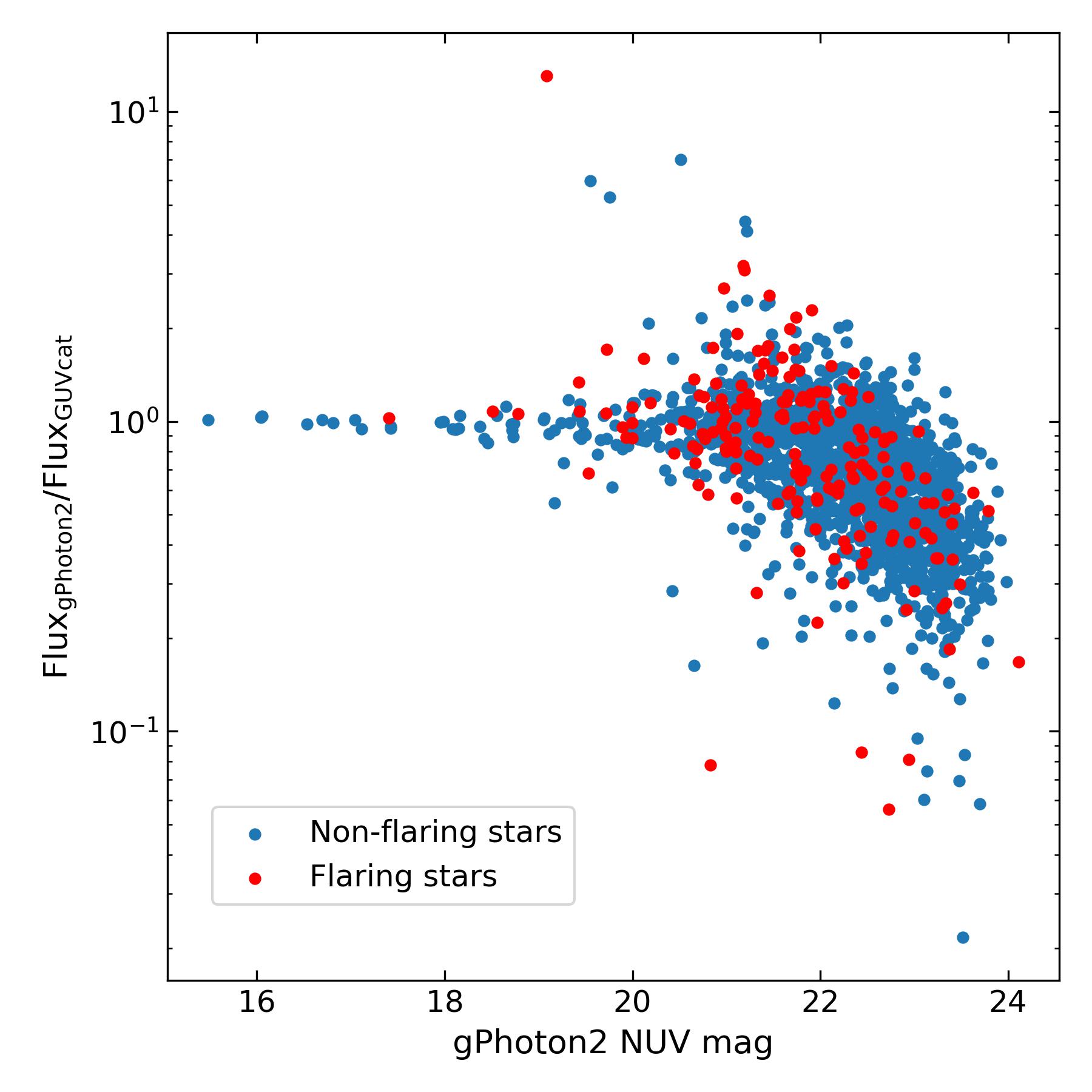}
    \caption{Comparison of GALEX NUV fluxes obtained from our data reduction with GUVcat as a function of NUV magnitude. GUVcat tends to overestimate fluxes of faint sources for both flaring and non-flaring stars as compared to \gphoton2.}
    \label{fig: gphoton-guvcat comparison}
\end{figure}

We compare the quiescent fluxes we obtain after masking all suspect and anomalous regions from GALEX lightcurves with the Galex UV Unique Source Catalog cross-matched with Gaia DR2  \citep[GUVcat; ][]{bianchi_revised_2017, bianchi_matched_2020} in Figure \ref{fig: gphoton-guvcat comparison}. This serves as a check for our data reduction process and the calibration of \gphoton2. We see good agreement between the two down to $\sim 20$th magnitude, beyond which GUVcat tends to overestimate fluxes as compared to gPhoton2. While this does suggest a systematic effect in either or possibly both pipelines, the difference is ultimately within the combined errors of the two measurements and hence not very significant.

\section{Estimating M-dwarf populations} \label{appendix: G-M estimation}

\begin{figure}
    \centering
    \includegraphics[width=\columnwidth]{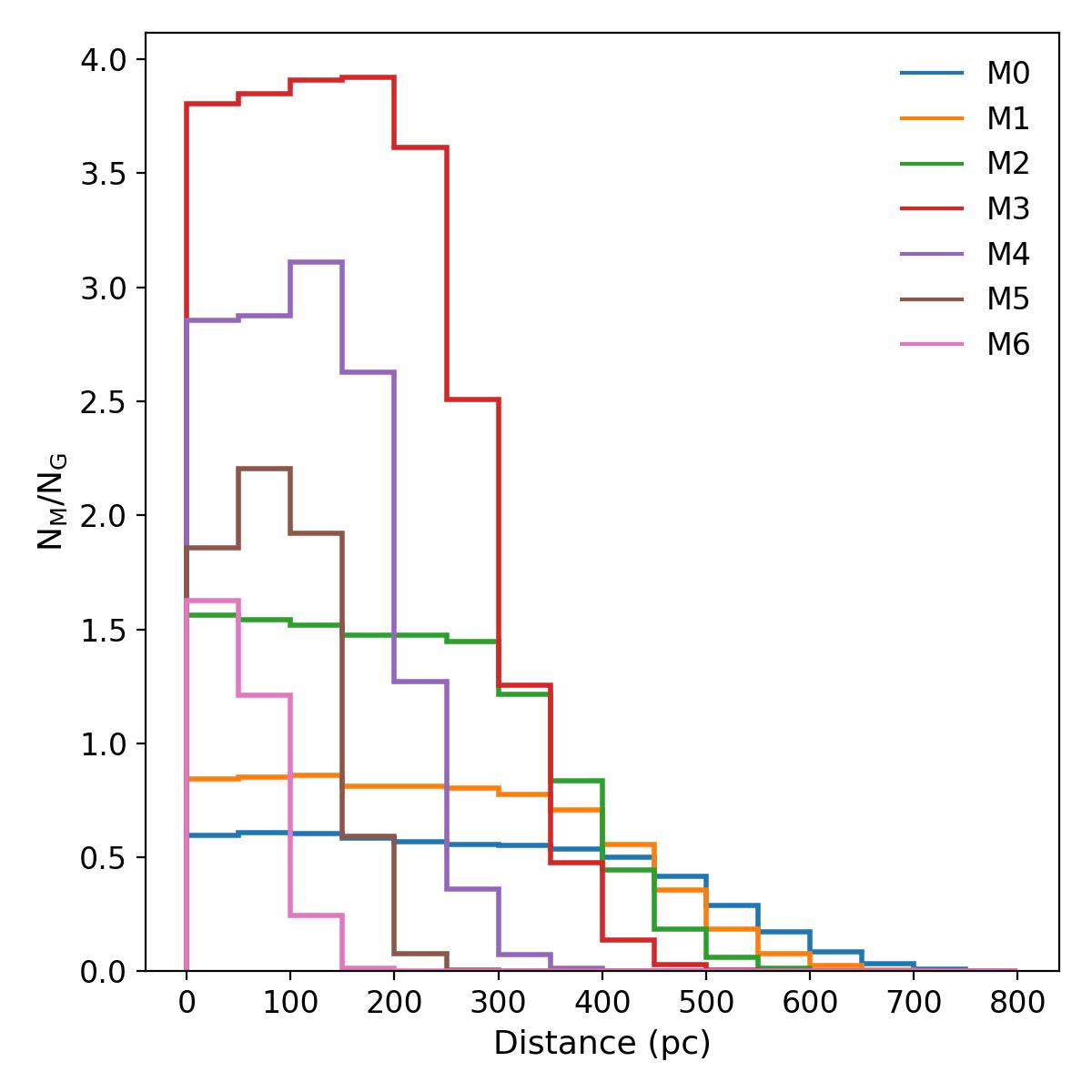}
    \caption{Ratio of numbers of M-Dwarfs to G-Dwarfs as a function of distance from Earth for galactic latitudes |b|>30\degree. }
    \label{fig:mbyg}
\end{figure}

We estimate the M-dwarf population till 800\,pc by tying the number density of M-dwarfs to that of G-dwarfs.
To begin, we obtain the observed G-dwarf sample upto 800 pc from the TIC in addition the M-dwarf sample obtained in section \ref{sec: Dataset}. G-dwarfs are selected as objects in the TIC with effective temperature between 5300\,K and 6000\,K and luminosity class `\noun{dwarf}'. Our procedure is as follows: We assume the ratio of number density of M-dwarfs to G-dwarfs as a function of M-dwarf spectral type is constant throughout space. We compute this ratio over the entire sky as a function of distance, excluding the regions near the galactic equator (|b|<30\degree) where the TIC may be affected due to extinction and overcrowding; \ultrasat{} will conduct its high cadence survey exclusively at galactic latitudes higher than $30\degree$. At higher latitudes the G-dwarf sample is presumed to be complete till $800\,$pc due to their much higher luminosities\footnote{At 800\,pc a G9 dwarf has Gaia G magnitude $\sim 15$, for which Gaia is essentially complete.}.  We then consider the peak of the density ratio distribution for each M-dwarf spectral type (Figure \ref{fig:mbyg}) to be the true value of the ratio. Thus the number of M-dwarfs of a given spectral type in any volume of space would be the number of G-dwarfs multiplied by this ratio. An important caveat here is that as the color-temperature relation used by the TIC is invalid below 3000\,K \citep{stassun_revised_2019}, the M5 and M6 stellar sample is incomplete even at short distances; and stars later than M6 are not included in the TIC. Hence our population estimates for stars cooler than M4 dwarfs are a lower limit.

We obtain absolute Gaia G-band magnitudes for M0 to M6 stars from \cite{pecaut_intrinsic_2013} and use the \ultrasat-Gaia G colors for M-dwarfs given by \cite{shvartzvald_ultrasat_2023} to obtain absolute \ultrasat{} magnitudes for each spectral type. We estimate  the nominal \ultrasat{} northern hemisphere pointing to be centered at the equatorial coordinate (220\degree,66\degree), corresponding to a galactic latitude of $47.6\degree$. We extract the observed distance distribution of G-dwarfs in the sky region between b=40\degree and b=55\degree, binned in steps of 25 pc, and multiply it by the number density ratio for each spectral type to derive their estimated number distribution. The population is then normalized to the \ultrasat{} FoV area of $170~\text{deg}^2$. In simulations, we assume the numbers of stars within each 25 pc bin to follow Poisson statistics, and the probability distribution for distances within a given bin to follow a power-law with index 3, as number density within the bin volume is approximately constant.

\clearpage
\onecolumngrid{
\section{Data Tables and Plots}

Table \ref{tab: startable} contains information on all stars in the \galex{} sample and Table \ref{tab: flaretable} lists all the visually verified flares in the sample. Tables \ref{tab: startable xmm} and \ref{tab: flaretable xmm} list all stars and detected flares in the \xmmom{} sample, respectively. Figures \ref{fig: eqd indiv fits} and \ref{fig: energy indiv fits} show observed FFDs in equivalent duration and energy for individual spectral types along with power-law fits.

\begin{table*}[h!p!]
\centering
\begin{tabularx}{0.87\textwidth}{cccccccccc}
\hline
TIC ID & N\textsubscript{flares} & Exposure Time & NUV Flux & $\Delta$ NUV Flux & M\textsubscript{NUV} & $\Delta$ M\textsubscript{NUV} & T\textsubscript{eff} & Dist & $\Delta$ Dist \\
 &  & $\mathrm{s}$ & $\mathrm{ct\,s^{-1}}$ & $\mathrm{ct\,s^{-1}}$ & $\mathrm{mag}$ & $\mathrm{mag}$ & $\mathrm{K}$ & $\mathrm{pc}$ & $\mathrm{pc}$ \\ \hline
818309 & 0 & 1600 & 0.099 & 0.011 & 17.234 & 0.123 & 3652 & 117.82 & 0.38 \\
856175 & 0 & 1400 & 0.032 & 0.004 & 18.238 & 0.137 & 3583 & 130.66 & 0.47 \\
893123 & 1 & 2560 & 0.595 & 0.020 & 17.944 & 0.037 & 3420 & 34.67 & 0.025 \\
916910 & 0 & 1600 & 0.060 & 0.007 & 17.985 & 0.130 & 3301 & 107.40 & 0.63 \\
921601 & 0 & 1300 & 0.068 & 0.013 & 17.292 & 0.203 & 3171 & 138.60 & 1.4 \\
970375 & 0 & 4280 & 0.271 & 0.012 & 15.940 & 0.049 & 3320 & 129.32 & 0.74 \\
972579 & 0 & 1560 & 0.034 & 0.010 & 19.073 & 0.311 & 3690 & 86.27 & 0.93 \\
1025507 & 0 & 1560 & 0.178 & 0.014 & 16.725 & 0.088 & 3587 & 111.21 & 0.35 \\
1029764 & 0 & 1560 & 0.162 & 0.012 & 16.110 & 0.081 & 3762 & 154.65 & 0.59 \\
1045737 & 0 & 13050 & 0.062 & 0.003 & 18.360 & 0.061 & 3741 & 88.92 & 0.13 \\
\hline
\end{tabularx}
\caption{Top 10 rows of the list of all stars in the GALEX sample. Effective temperatures are obtained from the TIC, and distance estimates are obtained from \cite{bailer-jones_estimating_2021}.}
\label{tab: startable}
\end{table*}

\begin{table*}[h!p!]
\centering
\begin{tabularx}{0.86\textwidth}{lrrrccccr}
\hline
TIC ID & Flare Start & ED & $\Delta$ ED & Energy & $\Delta$ Energy & Complete? & Flare Length & ${F_p}/{F_q}$ \\
 & $\mathrm{s}$ & $\mathrm{s}$ & $\mathrm{s}$ & $\mathrm{erg}$ & $\mathrm{erg}$ &  & $\mathrm{s}$ &  \\ \hline
142215003$^\dagger$ & 789338719 & 132084 & 704 & 4.60E+34 & 3E+32 & Y & 980 & 528.3 \\
73540267 & 990982754 & 270000 & 113662 & 6.08E+33 & 1E+32 & Y & 1100 & 994.3 \\
325273691$^\dagger$ & 766525751 & 63698 & 321 & 4.81E+33 & 1E+31 & N & 830 & 204.0 \\
393451028 & 760180732 & 26018 & 487 & 2.73E+33 & 8E+31 & N & 1520 & 83.4 \\
394580039 & 994874692 & 102501 & 1491 & 2.16E+33 & 6E+31 & Y & 1260 & 283.8 \\
117643740 & 817562647 & 40033 & 4951 & 1.54E+33 & 9E+31 & N & 1190 & 114.1 \\
287958625 & 926853917 & 35839 & 878 & 1.42E+33 & 2E+31 & N & 1620 & 57.4 \\
20963051 & 1006123987 & 13997 & 943 & 1.40E+33 & 4E+31 & Y & 1040 & 39.8 \\
9070340 & 969818263 & 17417 & 544 & 1.33E+33 & 1E+32 & N & 1540 & 21.3 \\
453516198 & 993803808 & 1787 & 41 & 1.11E+33 & 2E+31 & N & 290 & 16.3 \\
\hline
\end{tabularx}
\caption{Top 10 most energetic flares from the list of all flares detected in the \galex{} sample. ED refers to Equivalent Duration; ${F_p}/{F_q}$ is the ratio of peak flare to quiescent flux. Flare start times are in GALEX time = t\textsubscript{UNIX} - 315964800\,s.\\
$^\dagger$ Flare has reached non-linearity regime of the detector, hence energy and equivalent duration are lower limits.}
\label{tab: flaretable}
\end{table*}

\begin{table*}
\centering
\begin{tabularx}{0.76\textwidth}{cccccccc}
\hline
TIC ID & N\textsubscript{flares} & Exposure Time & UVW1 Flux & $\Delta$ UVW1 Flux & T\textsubscript{eff} & Dist & $\Delta$ Dist \\
 &  & $\mathrm{s}$ & $\mathrm{ct\,s^{-1}}$ & $\mathrm{ct\,s^{-1}}$ & $\mathrm{K}$ & $\mathrm{pc}$ & $\mathrm{pc}$ \\ \hline
2468648 & 2 & 33520 & 0.4160 & 0.0001 & 3377 & 27.64 & 0.019 \\
4750626 & 0 & 20560 & 0.3180 & 0.0002 & 3153 & 11.87 & 0.027 \\
5656273 & 9 & 100000 & 17.4011 & 0.0003 & 3815 & 24.92 & 0.015 \\
16909043 & 0 & 1220 & 5.2463 & 0.0090 & 3237 & 5.01 & 0.00093 \\
19028197 & 0 & 4420 & 2.0340 & 0.0012 & 3552 & 29.38 & 0.019 \\
71011069 & 0 & 4020 & 1.9544 & 0.0013 & 3723 & 46.79 & 0.062 \\
138819293 & 0 & 51700 & 8.2411 & 0.0003 & 3456 & 9.77 & 0.0029 \\
142206123 & 8 & 30500 & 104.2556 & 0.0019 & 3788 & 11.54 & 0.005 \\
144654557 & 0 & 42060 & 0.4900 & 0.0001 & 3166 & 14.56 & 0.0034 \\
159376971 & 0 & 3520 & 1.0236 & 0.0011 & 3880 & 67.08 & 0.049 \\
172278724 & 0 & 15300 & 0.9470 & 0.0003 & 3247 & 10.08 & 0.0044 \\
187391650 & 0 & 10480 & 0.4840 & 0.0003 & 3224 & 13.50 & 0.0086 \\
206402318 & 0 & 7520 & 0.1490 & 0.0003 & 3125 & 12.25 & 0.0029 \\
259901346 & 0 & 10840 & 0.0737 & 0.0001 & 3029 & 12.75 & 0.002 \\
259999047 & 1 & 49560 & 9.1139 & 0.0007 & 3409 & 9.60 & 0.0014 \\
269891011 & 0 & 3020 & 0.6191 & 0.0010 & 3449 & 38.36 & 0.071 \\
297812150 & 0 & 11600 & 1.2155 & 0.0004 & 3196 & 13.06 & 0.0057 \\
441803471 & 0 & 2220 & 0.1961 & 0.0010 & 3135 & 15.56 & 0.0072 \\
\hline
\end{tabularx}
\caption{List of all stars in the \xmmom{} sample. Effective temperatures are obtained from the TIC, and distance estimates are obtained from \cite{bailer-jones_estimating_2021}.}
\label{tab: startable xmm}
\end{table*}

\begin{table*}[h!p!]
\centering
\begin{tabularx}{0.82\textwidth}{lcrrccccr}
\hline
TIC ID & Flare Start & ED & $\Delta$ ED & Energy & $\Delta$ Energy & Complete? & Length & ${F_p}/{F_q}$ \\
 & MJD & $\mathrm{s}$ & $\mathrm{s}$ & $\mathrm{erg}$ & $\mathrm{erg}$ &  & $\mathrm{s}$ &  \\ \hline
2468648 & 58574.100657941 & 945 & 63 & 3.32E+30 & 2.20E+29 & Y & 220 & 25.0 \\
2468648 & 58573.710129173 & 3561 & 378 & 6.54E+30 & 2.67E+29 & Y & 20 & 356.3 \\
5656273 & 53508.054828485 & 21 & 2 & 2.39E+30 & 4.24E+29 & Y & 50 & 0.8 \\
5656273 & 53508.750009838 & 21 & 2 & 2.43E+30 & 3.85E+29 & Y & 40 & 0.9 \\
5656273 & 53508.704870949 & 33 & 3 & 3.71E+30 & 5.05E+29 & Y & 70 & 0.7 \\
5656273 & 53508.023615083 & 41 & 3 & 4.67E+30 & 6.52E+29 & Y & 120 & 0.8 \\
5656273 & 53508.642209534 & 45 & 3 & 5.02E+30 & 6.28E+29 & Y & 110 & 0.5 \\
5656273 & 53508.068485892 & 47 & 3 & 5.24E+30 & 7.27E+29 & Y & 150 & 0.6 \\
5656273 & 53508.031680336 & 101 & 5 & 1.13E+31 & 8.16E+29 & Y & 180 & 1.3 \\
5656273 & 53508.638390089 & 122 & 6 & 1.37E+31 & 7.65E+29 & Y & 150 & 2.1 \\
5656273 & 53508.710310764 & 167 & 7 & 1.86E+31 & 1.07E+30 & Y & 310 & 1.0 \\
142206123 & 52859.506752381 & 20 & 1 & 3.32E+30 & 2.48E+29 & Y & 160 & 0.2 \\
142206123 & 52859.310224750 & 24 & 1 & 3.62E+30 & 2.69E+29 & Y & 170 & 0.2 \\
142206123 & 52859.535226852 & 26 & 1 & 3.98E+30 & 2.52E+29 & Y & 140 & 0.7 \\
142206123 & 52859.344415367 & 31 & 1 & 4.60E+30 & 2.50E+29 & Y & 130 & 0.6 \\
142206123 & 52859.611718314 & 56 & 2 & 7.84E+30 & 3.39E+29 & N & 260 & 0.4 \\
142206123 & 52859.443816443 & 100 & 2 & 1.42E+31 & 3.73E+29 & Y & 250 & 0.7 \\
142206123 & 52859.477551105 & 301 & 4 & 4.77E+31 & 7.43E+29 & N & 760 & 0.9 \\
142206123 & 52859.688882994 & 588 & 6 & 8.41E+31 & 9.31E+29 & N & 1200 & 1.0 \\
259999047 & 58473.330244405 & 39 & 4 & 3.40E+29 & 4.48E+28 & Y & 130 & 0.6 \\
\hline
\end{tabularx}
\caption{List of all flares detected in the \xmmom{} sample. ED refers to Equivalent Duration; ${F_p}/{F_q}$ is the ratio of peak flare to quiescent flux.}
\label{tab: flaretable xmm}
\end{table*}

\begin{figure*}[h!p!]
    \centering
    \includegraphics[height=\textheight]{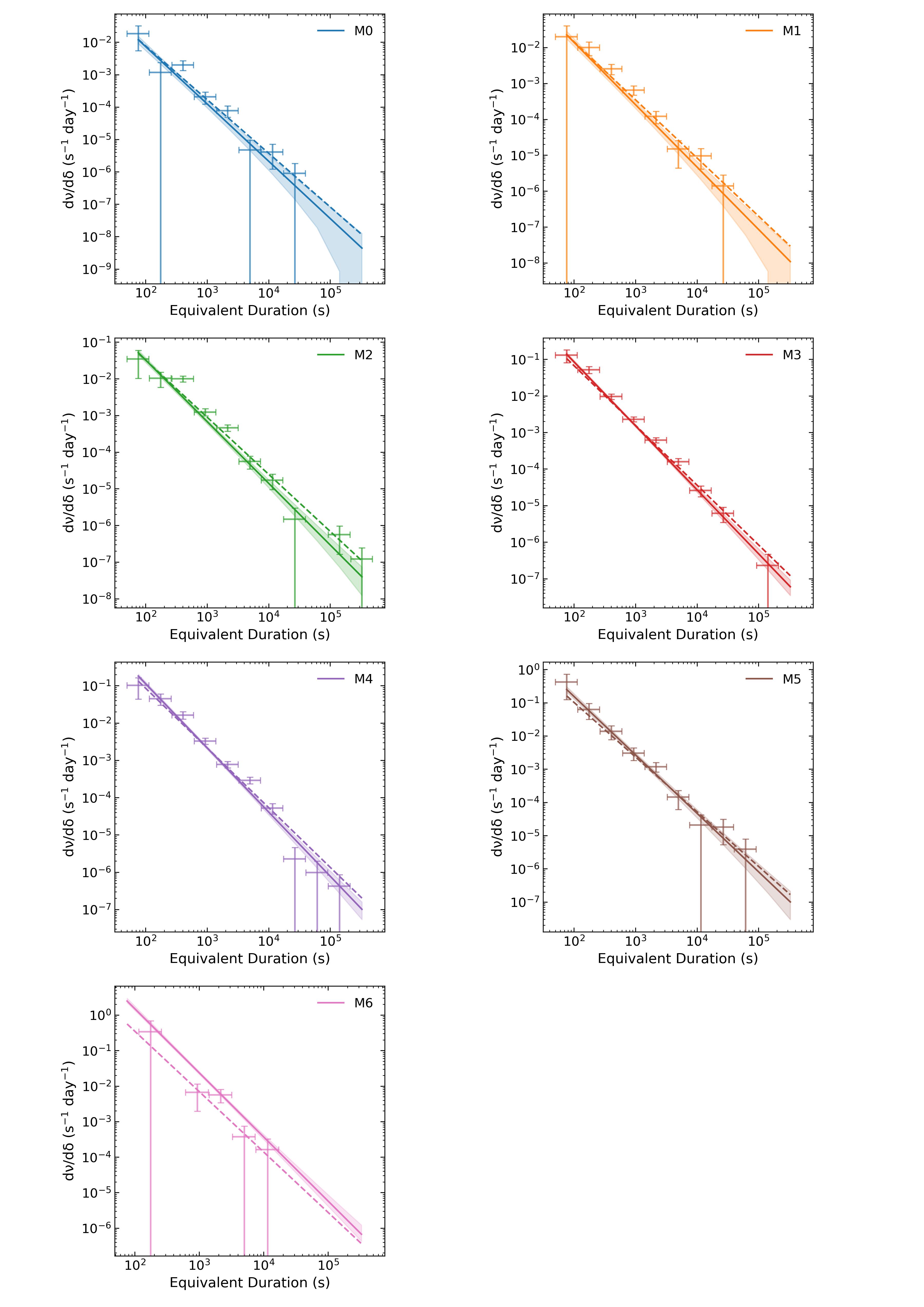}
    \caption{(a) GALEX equivalent duration flare frequency distributions and power-law fits for individual spectral types. Plot elements are the same as in Figure \ref{subfig: grouped ffd delta}.}
    \label{fig: eqd indiv fits}
\end{figure*}

\begin{figure*}[h!p!]
    \centering
    \includegraphics[height=\textheight]{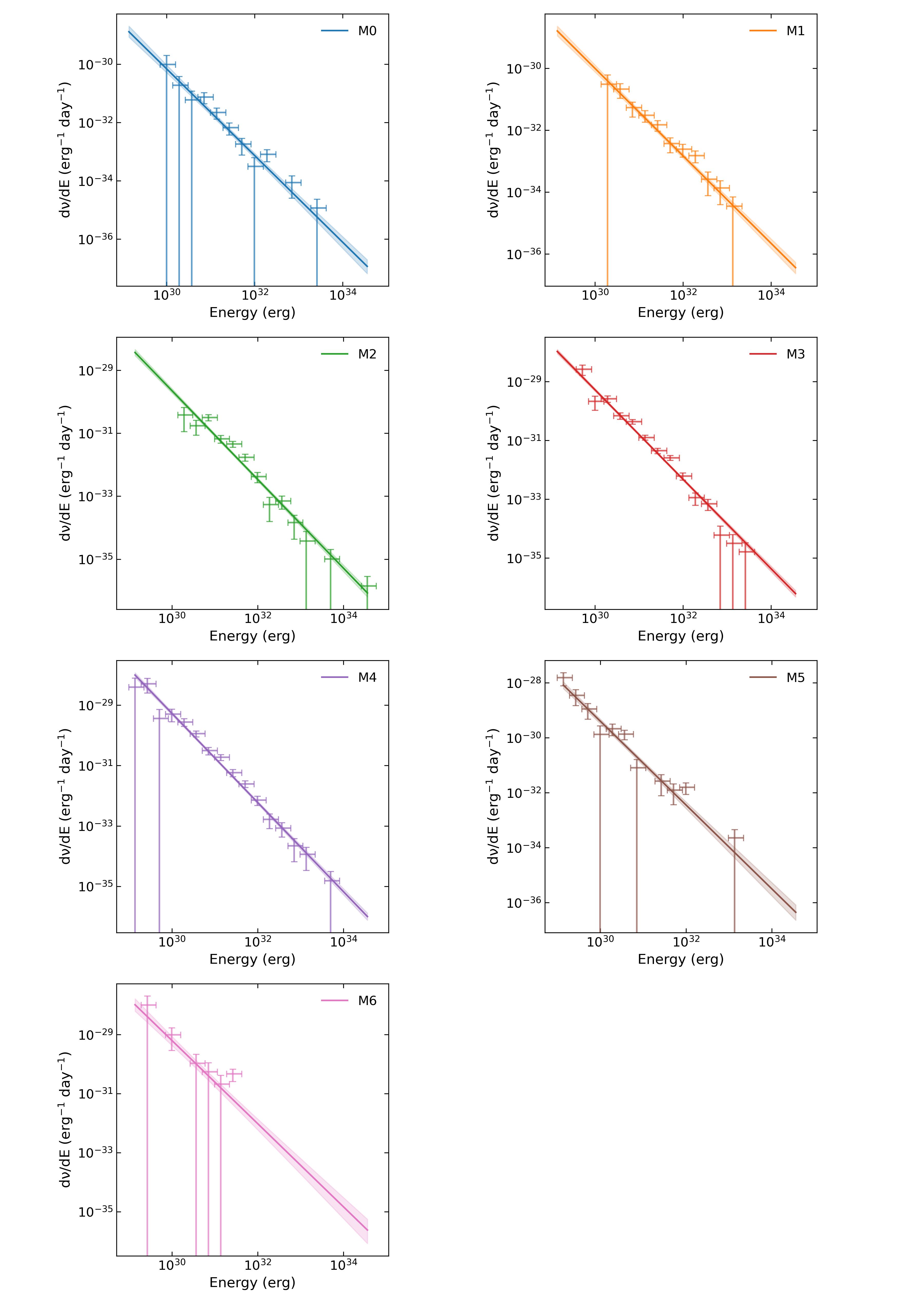}
    \caption{(a) GALEX energy flare frequency distributions and power-law fits for individual spectral types. Plot elements are the same as in Figure \ref{subfig: grouped ffd energy}.}
    \label{fig: energy indiv fits}
\end{figure*}

}

\newpage

\bibliography{main}
\bibliographystyle{aasjournal}

\end{document}